\documentclass{article}

\usepackage{arxiv}

\usepackage[utf8]{inputenc}  
\usepackage[T1]{fontenc}     
\usepackage{hyperref}        
\usepackage{url}             
\usepackage{booktabs}        
\usepackage{amsfonts}        
\usepackage{nicefrac}        
\usepackage{microtype}       
\usepackage{graphicx}
\usepackage{natbib}
\usepackage{doi}
\usepackage[svgnames]{xcolor}
\usepackage{listings}
\usepackage{thumbpdf,lmodern}
\usepackage{amsmath}
\usepackage{amssymb}
\usepackage{marvosym}

\DeclareMathOperator*{\argmin}{arg\,min}
\DeclareMathOperator*{\argmax}{arg\,max}

\title{\pkg{SEMgraph}: An \proglang{R} Package for Causal Network Inference of High-Throughput Data with Structural Equation Models}

\author{
	\myhref[DarkBlue]{https://orcid.org/0000-0003-0861-5512}{\includegraphics[scale=0.08]{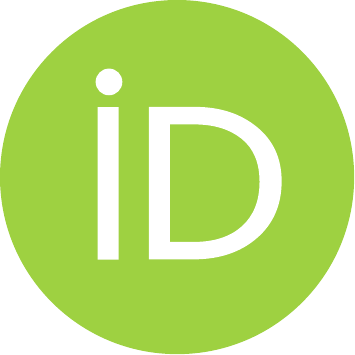}\hspace{1mm}Fernando Palluzzi}
	\vspace{1mm}
	\thanks{Corresponding author.} \\
	Department of Brain and Behavioral Sciences\\
	Universit\`a di Pavia\\
	27100 Pavia (PV), Italy \\
	\texttt{fernando.palluzzi@gmail.com} \\
	\And
	\myhref[DarkBlue]{https://orcid.org/0000-0001-6612-8566}{\includegraphics[scale=0.08]{orcid.pdf}\hspace{1mm}Mario Grassi}
	\vspace{1mm}\\
	Department of Brain and Behavioral Sciences\\
	Universit\`a di Pavia\\
	27100 Pavia (PV), Italy \\
	\texttt{mario.grassi@unipv.it} \\
}

\renewcommand{\headeright}{Preprint}
\renewcommand{\undertitle}{}
\renewcommand{\shorttitle}{Network Analysis and Causal Learning with SEMgraph}

\let\proglang=\textsf
\newcommand{\pkg}[1]{{\fontseries{b}\selectfont #1}}
\let\proglang=\textsf

\hypersetup{
pdftitle={Network Analysis and Causal Learning with SEMgraph},
pdfsubject={stat.AP, q-bio.MN},
pdfauthor={Fernando Palluzzi, Mario Grassi},
pdfkeywords={structural equation modeling, network analysis, causal inference, genomics, R},
colorlinks=true,
linkcolor=red,
filecolor=cyan,
urlcolor=DarkRed,
citecolor=DarkBlue,
}
\newcommand{\myhref}[3][DarkRed]{\href{#2}{\color{#1}{#3}}}

\begin{document}
\maketitle

\begin{abstract}

  With the advent of high-throughput sequencing (HTS) in molecular biology 
  and medicine, the need for scalable statistical solutions for modeling 
  complex biological systems has become of critical importance. The 
  increasing number of platforms and possible experimental scenarios 
  raised the problem of integrating large amounts of new heterogeneous 
  data and current knowledge, to test novel hypotheses and improve our 
  comprehension of physiological processes and diseases. 
  Although network theory provided a framework to represent biological 
  systems and study their hidden properties, different algorithms still 
  offer low reproducibility and robustness, dependence on user-defined 
  setup, and poor interpretability. Here we discuss the \proglang{R} 
  package \pkg{SEMgraph}, combining network analysis and causal inference 
  within the framework of structural equation modeling (SEM). It provides 
  a fully automated toolkit, managing complex biological systems as 
  multivariate networks, ensuring robustness and reproducibility 
  through data-driven evaluation of model architecture and perturbation, 
  that is readily interpretable in terms of causal effects among system 
  components. In addition, \pkg{SEMgraph} offers several functions for 
  perturbed path finding, model reduction, and parallelization options 
  for the analysis of large interaction networks.

\end{abstract}

\keywords{structural equation modeling \and network analysis \and causal inference \and causal learning \and R}

\section{Introduction}
Discovering and understanding the mechanisms underlying complex phenotypical 
traits is of primary importance in bio-medical research. A deeper and detailed 
knowledge of the physio-pathological events leading to the onset and 
progression of a disease enables a clearer estimation of disease risk and more 
accurate diagnosis, prognosis evaluation, and decision making, including 
treatment choice \citep{Interactions2015}. With the advent of the 
high-throughput sequencing (HTS) technologies, the actual complexity behind 
diseased (and generally, phenotypical) traits became prominent, opening 
up to the big data era also in molecular biology and medicine 
\citep{Sequencing2012}. Biological systems complexity arises from the 
interactions and reactions among their components (e.g., genomic elements, 
epigenomic modifications, DNA-binding proteins, miRNAs, receptors, signaling 
molecules) and the layered modularity of their compartments (e.g., cellular 
components, tissues, organs). Predicting the behavior of these components 
after external perturbation or intrinsic variability (e.g., genetic 
polymorphisms), is key for the discovery and prediction of disease-associated 
processes \citep{Network2020,Interactions2015,Sequencing2012}. 
Biological models are commonly represented by signaling pathways, chains of 
metabolic reactions, disease modules, or very large protein-protein interaction 
networks (also called interactomes) \citep{Interactions2015,NetworkMed2011}. 
Given the vast amount of publicly available bio-medical databases, the 
access to curated biological models is no longer a limitation. The key 
feature of these databases is the availability of structured bio-chemical 
and bio-medical information that can be readily converted into networks 
and statistical models: we generally refer to them as knowledge-based 
models (KBMs). KBMs provide a basis and a gold standard to improve 
exploratory methods, with some critical issues \citep{Interactions2015}. 
Fistly, defining a set of rules to convert a KBM to a causal model is key 
to test specific biological hypotheses and mechanisms, but it is not always 
trivial due to missing information 
or evidence level (e.g., experimental evidence versus inference from 
similarity or electronic annotation). 
Secondly, a KBM reflects current knowledge, constantly challenged by new 
experimental data that may reveal novel interactions and pathways. 
Finally, there must be clear statistical criteria to evaluate the initial 
causal model, reflecting biological properties of the system, and 
improving both model descriptive and predictive power 
\citep{Network2020,Interactions2015}.
Starting from current knowledge, network models should be updated and 
tested in a simple and clear workflow. From the computational point of 
view, the challenge is to free the user from chosing the initial setup, 
estimating algorithm and model parameters directly from quantitative 
data, with eﬃcient and parallelizable methods.\\
Motivated by this challenge, we developed the \proglang{R} package 
\pkg{SEMgraph}, based on structural equation modeling (SEM) \citep{SEM1989}, 
enabling causal inference on complex biological networks.
SEM are now a popular tool in causal inference \citep{Causality2009}, 
causal structure learning \citep{Causality2000}, and biostatistics. 
Path diagrams, often represented as acyclic mixed graphs, provide a 
backbone for model learning, data-driven model refinement and causal 
inference and discovery. 
HTS data is often structured into pathways or large networks, enabling 
either confirmatory or exploratory analysis of salient biological 
properties. Within \pkg{SEMgraph}, this is practically achieved through 
algorithm-assisted search for the optimal trade-off between best model 
fitting (i.e., the optimal context) and perturbation (i.e., exogenous 
influence) given data, in which knowledge is used as supplementary 
confirmatory information. 
In \pkg{SEMgraph}, the input network and the underlying statistical model 
are interchangeable representations of the same object: a set of interacting 
variables linked by causal relationships. This dual representation is 
opportunely manipulated to generate the final causal model, through a series 
of intermediate steps, including causal backbone estimation, adjustement 
of hidden confounding variables, graph extension, and model refinement 
to improve fitting, whith scalable solutions for large graphs.
In this work, we expose the relevant \pkg{SEMgraph} functions with examples 
of typical applications in genomics.\\
\pkg{SEMgraph} package is available under the GNU General Public License 
version 3 or higher (GPL $\geq$ 3) from CRAN repository,
and the latest stable version can be installed via:
\begin{lstlisting}
R> install.packages("SEMgraph")
\end{lstlisting}
The development version of \pkg{SEMgraph} can be installed from the GitHub 
repository, at \url{https://github.com/fernandoPalluzzi/SEMgraph}
through \pkg{devtools}:
\begin{lstlisting}
R> devtools::install_github(c("fernandoPalluzzi/SEMgraph"))
\end{lstlisting}

\section{Structural equation models} \label{sec:sem}

\subsection{SEM basics} \label{sec:sembasics}

SEM is a statistical framework for causal inference based on 
multivariate linear regression equations, where the response variable in 
one regression equation may appear as a predictor in another equation 
\citep{SEM1989, Causality2016}. 
SEM may be formulated to explicity include latent unobserved variables, 
but here we consider a setup in which the latent variable have been 
marginalized out and represented in the model only implicitly through 
possible correlations among unobserved latent confounders \citep{SEM1998}.\\
A SEM, is based on a system of structural (i.e., linear regression) equations 
definig a \emph{path diagram}, represented as a graph $G = (V, E)$, where 
$V$ is the set of nodes (i.e., variables) and $E$ is the set of edges 
(i.e., connections). 
The set $E$ may include both directed edges $k \rightarrow j  \, \, 
\mathrm{if}  \, \, k \in \mathrm{pa}(j)$ and bidirected edges 
$k \leftrightarrow j  \,  \, \mathrm{if} \, \, k \in \mathrm{sib}(j)$, 
where the \emph{parent} set $\mathrm{pa}(j)$, and the \emph{siblings} 
set $\mathrm{sib}(j)$, determine the system of linear equations, 
as follows:
\begin{equation} \label{eq:semsys}
Y_j = \displaystyle\sum_{k  \, \in  \, \mathrm{pa}(j)}^{} 
\beta_{jk} Y_k + U_j \qquad j \in V
\end{equation}
\begin{equation} \label{eq:semcov}
\mathrm{cov}(U_j; U_k) =
\begin{cases}
\psi_{jk} & \quad \mathrm{if} \, j = k \mathrm{ \, \, or \, \, } k \in 
\mathrm{sib}(j)\\
0 & \quad \text{otherwise}
\end{cases}
\end{equation}
where $Y_j$  and $U_j$ are an observed variable and  an unobserved error term,
respectively; $\beta_{jk}$ are regression coefficients,  and a covariance $\psi_{jk}$
indicates that errors are dependent, which is assumed when there exists an
unobserved  (i.e.latent) confounder between $k$ and $j$.\\
A path diagram is also a formal tool to evaluate the hierarchical 
structure of a system, where we can identify \emph{exogenous variables} 
as system elements with empty parents set, and \emph{endogenous variables}, 
having at least one parent variable in at least one structural equation 
of the SEM. In graph theory, exogenous variables are \emph{source} nodes, 
with incoming connectivity equal to 0, whilst endogenous variables are 
nodes with non-zero incoming connectivity. Endogenous variables 
can be further divided into \emph{connectors}, with non-zero outgoing 
connectivity, and \emph{sinks}, having no outgoing connections. Given 
these notions, we consider three types of fundamental path diagrams to 
describe high-throughput data structure:
\begin{itemize}
\item Directed Acyclic Graphs (DAGs), composed by directed edges 
($k \rightarrow j$) only, whose magnitude is quantified through path 
coefficients $\beta_{jk}$, and all covariances are null (i.e., 
$\psi_{jk} = 0$). In addition, loops are not allowed in a DAG.
\item Bow-free Acyclic Paths (BAPs), having acyclic directed edges 
($k \rightarrow j$), and bidirected connections ($k \leftrightarrow j$) 
only if the $k$-th and $j$-th variable do not share any directed link 
(i.e., they are bow-free). As a consequence, in a BAP, if 
$\beta_{jk} \neq 0$ then $\psi_{jk} = 0$.
\item Covariance models, as a special case of BAP in which all 
$\beta_{jk} = 0$. Therefore, only covariances $\psi_{jk}$ may have 
non-zero values.
\end{itemize}
These three models are simple graphs; i.e., they have at most one edge 
between any pair of nodes, and are all identifiable, such that the parameter 
matrices $B$ and $\Psi$ can be uniquely estimated from the population 
covariance matrix of the observed variables for nearly every parameter 
choice \citep{SEM2002,SEM1998}.

\subsubsection{SEM fitting} \label{sec:semfit}

From the computational point of view, it is convenient to write 
Equations~\ref{eq:semsys} and \ref{eq:semcov} in matrix form as: 
$Y = BY + U$ and $\mathrm{cov}(U) = \Psi$.
Assuming random variables with zero mean vector ($\mu(\theta) = 0$), 
the covariance matrix of the joint distribution of $p$ variables 
$Y$ is given by:
\begin{equation} \label{eq:jointcov}
\Sigma(\theta) = (I - B)^{-1} \Psi(I - B)^{-T}
\end{equation}
where the set of free parameters $\theta = (\beta; \psi)$ has dimension 
$t$. $B$ is the path coefficient matrix, $\Psi$ is the covariance matrix, 
and $I$ is the identity matrix, all of them having dimension $p \times p$. 
Generally, in the SEM framework, free (i.e., unknown) parameters 
$\theta$ are computed by Maximum Likelihood Estimation (MLE), assuming 
all model variables as jointly gaussian, so that the estimated covariance 
matrix $\Sigma(\hat\theta)$ is close to the observed sample covariance 
matrix $S$. This is obtained by maximizing  (up to an additive constant) 
the model log-likelihood function $\mathrm{log}L(\theta)$ given data 
\citep[p. 135]{SEM1989}. 
\begin{equation} \label{eq:logl}
\argmax_{\theta \, \in \, \mathbb{R}^t} \, 
\mathrm{log}L(\theta) = -\frac{n}{2} (\mathrm{log} \, \mathrm{det} \, 
\Sigma(\theta) + \mathrm{tr}[\Sigma(\theta)^{-1} S])
\end{equation}
From the expected Fisher\textquotesingle s information matrix of the 
likelihood function, standard errors, $\mathrm{SE}(\hat\theta)$ of the 
MLE $\hat\theta$ are extracted. MLE approximates a normal distribution 
and the P-values are computed through the test statistic 
$z = \hat\theta / \mathrm{SE}(\hat\theta)$ with 95\% confidence intervals: 
$\hat\theta \pm 1.96 \, \mathrm{SE}(\hat\theta)$. An advantage of MLE is 
that its estimates are in general scale invariant and scale free 
\citep[p. 109]{SEM1989}. Therefore, the values of the fit function 
do not depend on whether correlation or covariance matrices are analyzed, 
and whether original or transformed data are used. Model assessment is 
based on a chi-squared likelihood ratio test (LRT) statistic, known as 
model deviance:
\begin{equation} \label{eq:chisqlrt}
\chi^2 = -2 \, \mathrm{log \, LRT} = 
-2 \, [ \, \mathrm{log}L(\hat\theta) - \mathrm{log}L(\theta_\mathrm{max}) \, ]
\end{equation}
where $\mathrm{log}L(\hat\theta)$ is the log-likelihood 
Equation~\ref{eq:logl} evaluated to model-implied covariance matrix, 
$\Sigma(\hat\theta)$ and $\mathrm{log}L(\theta_\mathrm{max})$ is the 
log-likelihood for an exact fit; i.e., $\Sigma(\hat\theta) = S$. 
P-values are derived either from the $\chi^2(\mathrm{df})$ distribution 
with $\mathrm{df} = p(p + 1)/2 - t$ degrees of freedom, or from a resampling 
bootstrap distribution \citep{Bootstrap1992}. Non-significant P-values 
(P > 0.05) indicate that the model provides a good fit to data (i.e., 
the elements of $S - \Sigma(\hat\theta)$, should be close to zero). 
Alternatively to the chi-square test, the Akaike\textquotesingle s 
information criterion (AIC) \citep{AIC1974} can be used to compare fitted 
to saturated model, defined in SEM as \citep{EQS6}:
\begin{equation} \label{eq:aic}
\mathrm{AIC} \, = \, -2 \, \mathrm{log} L(\hat\theta) + 2t \, \approx \, 
\chi^2 - 2 \mathrm{df}
\end{equation}
where the rightmost member in Equation~\ref{eq:aic} is equal to the left 
member minus the constant term $p(p + 1)/2$. The model with the minimum 
AIC value is regarded as the best fitting model. In the chi-square (or 
deviance) metric it has been suggested that a ratio between the magnitude 
of $\chi^2$ and the expected value of the sample distribution 
$E(\chi^2) = \mathrm{df}$ less than 2 and between 2 and 3 is indicative 
of a good and acceptable data-model fit, respectively \citep{SEM2003}. 
The relationship between AIC and $\chi^2/\mathrm{df}$ thresholds become 
more evident by comparing Equation~\ref{eq:chisqlrt} and 
Equation~\ref{eq:aic}. For the saturated model $\mathrm{AIC} = 0$, and 
the fitted model should be selected if $\mathrm{AIC} < 0$, which is 
equivalent to the condition $\chi^2/\mathrm{df} < 2$.\\
Another approximate SEM fit index comparing two models (fitted vs. 
saturated) as the chi-square test (or the chi-square ratio), is the 
Standardized Root Mean-squared Residual (SRMR), an overall descriptive 
statistic based on all pairwise differences between observed sample 
covariances ($s$) and implied model covariances ($\sigma$):
\begin{equation} \label{eq:srmr}
\mathrm{SRMR} = \sqrt{\frac{\sum_{j = 1}^{p - 1} \sum_{k = j + 1}^{p} 
(s_{jk} - \sigma_{jk})^2 / s_{jj} s_{kk}}
{p (p + 1) / 2}}
\end{equation}
SRMR values range from 0 to 1, where 0 is equivalent to a perfect fit. 
The acceptable range for the SRMR index is between 0 and 0.08 
\citep{GOF1999}.\\
If the model is a DAG, a global fitting statistic, based on the directed 
separation (d-separation) concept, can be applied \citep{SEM2000}.
In a DAG, missing edges between  nodes imply a series of independence 
relationships between variables (either direct or indirect). These 
independences are implied by the topology of the DAG and are determined 
through d-separation: two nodes, $Y_j$ and $Y_k$, are d-separated by a 
set of nodes $S$ if conditioning on all members in $S$ blocks all 
confounding (or \emph{backdoor}) paths between $Y_j$ and $Y_k$ 
\citep{SEM1998, Causality1990}. 
In a DAG, with $Y_j$ having a higher causal order than $Y_k$, it is 
possible to find a minimal set of conditional independencies $B_U$ 
implying all the other possible independencies, defined by: 
$B_U = \left\{ Y_j \perp Y_k \, | \, \mathrm{pa}(j) \cup 
\mathrm{pa}(k) \mathrm{,} \, \, j > k \right\}$.
The number of conditional independence constraints in the basis set $B_U$ 
equals the number of missing edges, corresponding to the number of degrees 
of freedom (df) of the model. If the graph is not very large or very sparse, 
it is possible to perform local testing of all missing edges separately, 
using the Fisher\textquotesingle s z-transform of the partial correlation. 
An edge ($k$; $j$) is absent in the graph when the null hypothesis 
$\mathrm{H_0} \mathrm{: cor}(Y_j; Y_k \, | \, \mathrm{pa}(j) \cup 
\mathrm{pa}(k)) = 0$ is not rejected. These individual tests implied by 
the basis set $B_U$ are mutually independent, thus their P-values $p_r$ 
can be combined in an overall test of the fitted model (i.e., the DAG) 
using Fisher\textquotesingle s statistic: 
\begin{equation} \label{eq:fisherc}
C = -2 \sum_{r = 1}^{R} \mathrm{log}(p_b)
\end{equation}
This statistic follows a chi-squared distribution with 
$\mathrm{df} = 2\times(\mathrm{number \, of \, missing \, edges})$. 
A non-significant P-value (P > 0.05) of $C$ indicates that the model
provides a good fit to data.

\subsubsection{Decomposition of effects} \label{sec:effdec}

In observational studies, as in network biology and medicine, there is 
the need for assessing causality over paths (i.e., chains of direct 
effects $X \rightarrow \dots \rightarrow Y$) having biological relevance. 
One important feature of SEM is the decomposition of effects between 
variables. We may define three types of causal effects: direct effect (DE), 
indirect effect (IE), and total effect (TE). A DE is the causal effect 
$X \rightarrow Y$ of the $j$-th variable ($X$) on the $k$-th variable 
($Y$) of the model, when all other variables are kept constant (i.e., 
the effect quantified by path coefficients $\beta_{jk}$). Keeping the 
other variables constant will exclude all causal paths between $X$ and 
$Y$, with the exception of the direct connection $X \rightarrow Y$ 
\citep{SEM1998}; therefore the DE does not consider mediators effect. 
In a graph, a \emph{path} between two nodes $X$ and $Y$ can be viewed as 
a sequence of edges that may have either the same or different direction 
respect to neighbouring connections. A \emph{directed path} between two 
nodes is a sequence of edges with the same direction, where node $X$ is 
an \emph{ancestor} of $Y$, and $Y$ is a \emph{descendant} of $X$. The 
TE includes the contribution of all directed paths connecting $X$ and 
$Y$, whereas the IE can be defined as the difference 
$\mathrm{TE} - \mathrm{DE}$.\\
Let us consider an acyclic mixed graph $G$ (either a DAG or a BAP) and a 
directed path $\pi \in G$, traveling from node $X$ to node $Y$, having 
length (i.e., number of edges) equal to $r$. Every $j$-th directed edge 
in $\pi$ correspond to a DE quantified by a path coefficient $\beta_{j;j+1}$. 
The causal effect of $X$ on $Y$ through all the intermediate edges is 
given by the product of the underlying beta coefficients along a directed 
path from $X$ to $Y$.
In other words, we may consider $\pi$ as the path through which information 
is propagated from the source node $X$ to the target node $Y$. If there 
is more than one directed path $\pi_s(s = 1, ..., r(s))$ from $X$ to $Y$ 
in $G$, the TE will be the sum of the contribution of each alternative 
path $\pi$ through which information propagates from $X$ to $Y$:
\begin{equation} \label{eq:te}
\mathrm{TE} = \sum_{s}{} \pi_s = \sum_{s}^{} \prod_{j = 0}^{r(s)} 
\beta_{j;j+1}
\end{equation}
The nodes of an acyclic mixed graph can be ordered topologically, 
such that we observe a directed edge $j \rightarrow k$ only if $j < k$. 
All possible paths from $j$ to $k$ are given by $\left[ \, 
\sum_{r = 0}^{\infty} B^r \, \right]_{jk}$. 
Under node topological ordering, the path coefficents matrix $B$ is 
strictly lower-triangular, it is invertible, and 
$(I - B)^{-1} = I + B + B^2 + ...$, implying \citep{Identify2011}:
\begin{align} \label{eq:directEff}
\mathrm{TE}_{jk} &= (I - B)_{jk}^{-1}\\
\mathrm{DE}_{jk} &= B_{jk}\\
\mathrm{IE}_{jk} &= (I - B)_{jk}^{-1} - B_{jk}
\end{align}
Generally, in observational studies and genomics, the interaction between 
pairs of variables is estimated as the direct effect of the source variable 
$X$ on the target variable $Y$, when all other predictors are kept constant. 
However, this interpretation is incomplete for systems in which mediators 
effects is not negligible, as in case of perturbation propagation though 
nodes of a community or a signaling pathway. In these cases, the TE is a 
more appropriate estimation, considering the simultaneous variation of 
all mediators. A formal definition of TE, as average causal effect (ACE), 
is provided by the post-intervention $do$-calculus, defined in 
\cite{Causality2009}:

\begin{equation} \label{eq:ace}
\mathrm{ACE} = \mathrm{E}[Y \, | \, \mathrm{do}(X = x + 1)] - 
\mathrm{E}[Y \, | \, \mathrm{do}(X = x)]
\end{equation}

where $\mathrm{E}[Y \, | \, \mathrm{do}(X = x)]$ denotes the expected 
value of $Y$ when $X$ is fixed to a reference value $x$ by external 
intervention, as in a randomized experiment. In nonlinear models, the 
ACE will depend on the reference point. However, in a linear Gaussian SEM, 
$x$ can assume every arbitrary value and the intervention effect (or causal 
effect) will be a real-valued parameter, given by \citep{Causality2009}:
\begin{equation} \label{eq:ace}
\mathrm{ACE} = \frac{\partial}{\partial x} \mathrm{E}[Y \, | \, 
\mathrm{do}(X = x)]
\end{equation}
In acyclic mixed graphs, this constant parameter is given by the TE 
computed with the path method as $\mathrm{ACE}_{jk} = (I - B)_{jk}^{-1}$. 
Alternatively, when the causal model is a DAG, a simple way to compute 
the ACE is by applying Pearl\textquotesingle s backdoor criterion 
\citep{SEM1998}, allowing ACE estimation through regression. The parent 
set $\mathrm{pa}(X)$ of $X$ blocks all backdoor (i.e., confounding) paths 
from $X$ to $Y$, and the ACE is equal to the $\theta_{YX|Z}$ 
coefficient in a multiple regression of $Y$ on $X + \mathrm{pa}(X)$ 
\citep{Causality2009}. However, adjusting for $\mathrm{pa}(X)$ is typically 
inefficient with respect to its asymptotic variance, and an optimal 
adjustament set (O-set) with smallest asymptotic variance is obtained 
using the parent set of $Y$, $\mathrm{pa}(Y \, | \, D^{XY})$, in a suitable 
latent projection graph $D^{XY}$, called the forbidden projection 
\citep{oset2020}. The ACE is then computed as the $\theta_{YX|Z}$ 
coefficient in a multiple regression of $Y$ on 
$X + \mathrm{pa}(Y \, | \, D^{XY})$.

\subsection{Evaluating system perturbation with two-group SEM} \label{sec:sem2gr}

In several applications, the concept of \emph{perturbation} arises when 
a system is altered (i.e., changed) by one or more external influences 
affecting its \emph{behaviour} respect to a reference state (often described 
as \emph{physiological} or \emph{healthy}). However, in most cases, the 
mechanisms and extent of the alterations are unknown and data-driven discovery 
based on the comparison between experimental (i.e., altered) and healthy 
samples is the best possible option.\\
A possible approach to the evaluation of system perturbation is multigroup 
SEM \citep[p. 355]{SEM1989}. In \pkg{SEMgraph} a two-group SEM is implemented 
either using an exogenous group variable acting over a common model, or 
building a separate model for each group and comparing them. In the 
former, the experimental condition is compared to a control one through 
the use of an exogenous binary \emph{group} variable $X = \{0, 1\}$ acting 
on every node of the network. This model is converted to a system of linear 
equations that is common to both conditions, with $\mu(\theta) = 0$ and 
$\Sigma(\theta)$ being the implied mean vector and covariance matrix of 
the \emph{common model}:
\begin{align} \label{eq:semCommon}
Y_j &= \beta_{j} X + U_j \qquad \qquad \qquad \qquad 
\, \, \, \, \, j \in V(x)\\
Y_j &= \displaystyle\sum_{k  \, \in  \, \mathrm{pa}(j)}^{} 
\beta_{jk} Y_k + \beta_jX + U_j \qquad j \in V(y)
\end{align}
where $V(x)$ and $V(y)$ are the sets of exogenous (i.e., sources) and 
endogenous (i.e., connectors and sinks) variables, respectively. 
Coefficients $\beta_j$ (adjusted by the parents of the $j$-th node) 
determine the effect of the group on the $j$-th node, while the \emph{common} 
path coefficients $\beta_{jk}$ represent regression coefficients, adjusted 
by group effect. This type of SEM enables the identification of differentially 
regulated nodes (DRNs); i.e., variables showing a statistically significant 
variation in their activity (e.g., gene expression) in the experimental 
group respect to the control one. Alternatively, the two groups of samples 
(or subjects) are kept separated, with two different systems of linear 
equations:
\begin{equation} \label{eq:sem2mods1}
Y_j^{(1)} = \displaystyle\sum_{k  \, \in  \, \mathrm{pa}(j)}^{} 
\beta_{jk}^{(1)} Y_k^{(1)} + U_j^{(1)} \qquad j \in V(y)
\end{equation}
\begin{equation} \label{eq:sem2mods0}
Y_j^{(0)} = \displaystyle\sum_{k  \, \in  \, \mathrm{pa}(j)}^{} 
\beta_{jk}^{(0)} Y_k^{(0)} + U_j^{(0)} \qquad j \in V(y)
\end{equation}
This enables the identification of differentially regulated edges (DREs).
We define $\mu_1(\theta) = 0$ and $\Sigma_1(\theta)$ as 
the model-implied mean vector and covariance matrix for the experimental 
group (group 1), and $\mu_0(\theta) = 0$ and $\Sigma_0(\theta)$ the 
corresponding moments for the control group (group 0), respectively.
Perturbation tests in the common-model and two-models approaches are based 
on the definition of two different test statistics:
\begin{itemize}
\item $z_C = \beta_j / \mathrm{SE}(\beta_j)$, testing the null value 
for path coefficients $\beta_j$ of the group variable $X$ and evaluating 
node activation or inhibition;
\item $z_D = (\beta_{jk}^{(1)} - \beta_{jk}^{(0)}) / 
\mathrm{SE}(\beta_{jk}^{(1)} - \beta_{jk}^{(0)})$, testing the null value 
for path coefficients $\beta_{jk}$ differences between groups and evaluating 
edge activation or inhibition.
\end{itemize}
In both approaches, parameters are estimated through MLE and P-values for 
the $z$ statistics are derived asymptotically from the $N(0,1)$ standard 
Gaussian distribution. The descriptive overall group perturbation on either 
nodes or edges can be computed, for both node and edge differences, based 
on the Brown\textquotesingle s method for combining non independent, 
one-sided significance tests \citep{Brown1975}. The method computes the 
sum of one-sided pvalues: $X^2 = -2 \sum_{j}^{} \mathrm{log}(p_j)$, where 
the direction is chosen according to the alternative hypothesis 
($\mathrm{H_1}$), and the overall P-value is obtained from the chi-square 
distribution with new degrees of freedom $f$ and a correction factor $c$ 
to take into consideration the correlation among P-values \citep{Brown1975}.
The conversion of two-sided pvalues in one-sided pvalues is performed 
according to the sign of the z-test:
\begin{equation} \label{eq:H1plus}
\mathrm{H}_1 \text{: with at least one} \, \, \beta_j > 0 \implies p_j^{(+)} =
\begin{cases}
p_j / 2 & \quad \mathrm{if} \, z_j > 0\\
1 - p_j / 2 & \quad \mathrm{if} \, z_j < 0
\end{cases}
\end{equation}
\begin{equation} \label{eq:H1minus}
\mathrm{H}_1 \text{: with at least one} \, \, \beta_j < 0 \implies p_j^{(-)} =
\begin{cases}
p_j / 2 & \quad \mathrm{if} \, z_j < 0\\
1 - p_j / 2 & \quad \mathrm{if} \, z_j > 0
\end{cases}
\end{equation}
If the overall P-value < $\alpha$ (i.e., the significance level), we define 
node (or edge) perturbation as \emph{activated} when the direction of the 
alternative hypothesis is positive. Conversely, the status is \emph{inhibited} 
if the direction is negative. 

\subsection{Existing R packages for SEM} \label{sec:sempkgs}

There are many popular software packages for conducting SEM analysis, 
including commercial programs like \proglang{LISREL} \citep{LISREL10}, 
\proglang{EQS} \citep{EQS6}, and \proglang{Mplus} \citep{Mplus8}. Within 
the \proglang{R} environment \citep{R2020}, \pkg{lavaan} \citep{lavaan2012} 
is the most popular package for SEM and latent variable analysis, although 
alternative \proglang{R} packages are available, including: \pkg{sem} 
\citep{sem2006}, \pkg{OpenMx} \citep{OpenMx2011}, or \pkg{RAMpath} 
\citep{RAMpath2015}. All these packages use a specific model syntax or 
model matrix specification. The specialized package \pkg{dagitty} 
\citep{dagitty2016} estimates causal effects by covariate adjustment sets 
in four classes of causal models: DAGs, maximal ancestral graphs (MAGs), 
completed partially DAGs (CPDAGs), and partial ancestral graph (PAGs). 
Finally, \pkg{piecewiseSEM} \citep{piecewiseSEM2016} enables the analysis 
of linear, non-linear, mixed, and survival models as a SEM. With the 
availability of large genome-wide data sets, several existing \proglang{R} 
packages implemented SEM-based strategies for Genome-Wide Association 
Studies (GWAS). Package \pkg{GenomicSEM} \citep{GenomicSEM2019} uses SEM 
for modeling the multivariate genetic architecture of groups of correlated 
traits, incorporating the genetic covariance structure into a multivariate 
GWAS framework. Package \pkg{GW-SEM} \citep{GWSEM2017}, based on 
\pkg{OpenMx} \citep{OpenMx2011}, does SEM association analysis of SNPs 
with multiple phenotypes or latent constructs on a genome-wide basis.\\
Several recent SEM applications led to the development of sparse data 
analysis methods. Package \pkg{regsem} \citep{regsem2016}, designed 
for fitting common classes of SEM models with low dimensional data ($n > p$), 
uses \pkg{lavaan} outut for subsequent penalized likelihood analysis. 
Package \pkg{lslx} \citep{lslx2018} adopts a \pkg{lavaan}-like model syntax, 
where users can set each coefficient as free, fixed, or penalized. 
Finally, package \pkg{sparseSEM} \citep{SparseSEM2013} was developed for 
inferring gene regulatory networks from high-dimensional gene expression 
data and genetic makers.\\
Current SEM-based \proglang{R} packages and programming languages do not 
provide environments for automated and data-driven causal inference for 
network biology and medicine, integrating model syntax with graph analysis. 
With the adjectives \emph{automated} and \emph{data-driven}, we highlight 
the possibility to import, build, manage, and improve causal models directly 
leveraging on knowledge (i.e., the input graph), quantitative data, and 
a possible exogenous perturbation source (e.g., a phenotypical trait or 
a disease). Therefore, the \proglang{R} package \pkg{SEMgraph} comes with 
the following functionalities:
\begin{itemize}
\item Interchangeable model representation as either an \pkg{igraph} object 
or the corresponding SEM in \pkg{lavaan} syntax. Model management functions 
include automated covariance matrix regularization, graph-to-SEM or graph-to-DAG
conversion, and graph creation from correlation matrices.
\item Automated data-driven model building and improvement, through causal 
structure learning, bow-free interaction search, and latent variable 
confounding adjustment.
\item Perturbed paths finding, community searching, and sample scoring, 
together with graph plotting utilities, tracing model architecture 
modifications and perturbation (i.e., activation or repression) routes.
\item Heuristic graph filtering, node and edge weighting, resampling and 
parallelization settings for fast fitting in case of very large models.
\end{itemize}
This means letting the package finding possible solutions for 
high dimensionality, computational issues, and optimal causal 
architecture search.\\

\section{The SEMgraph package} \label{sec:semgraph}

\begin{figure}[t!]
\centering
\includegraphics[scale=0.7]{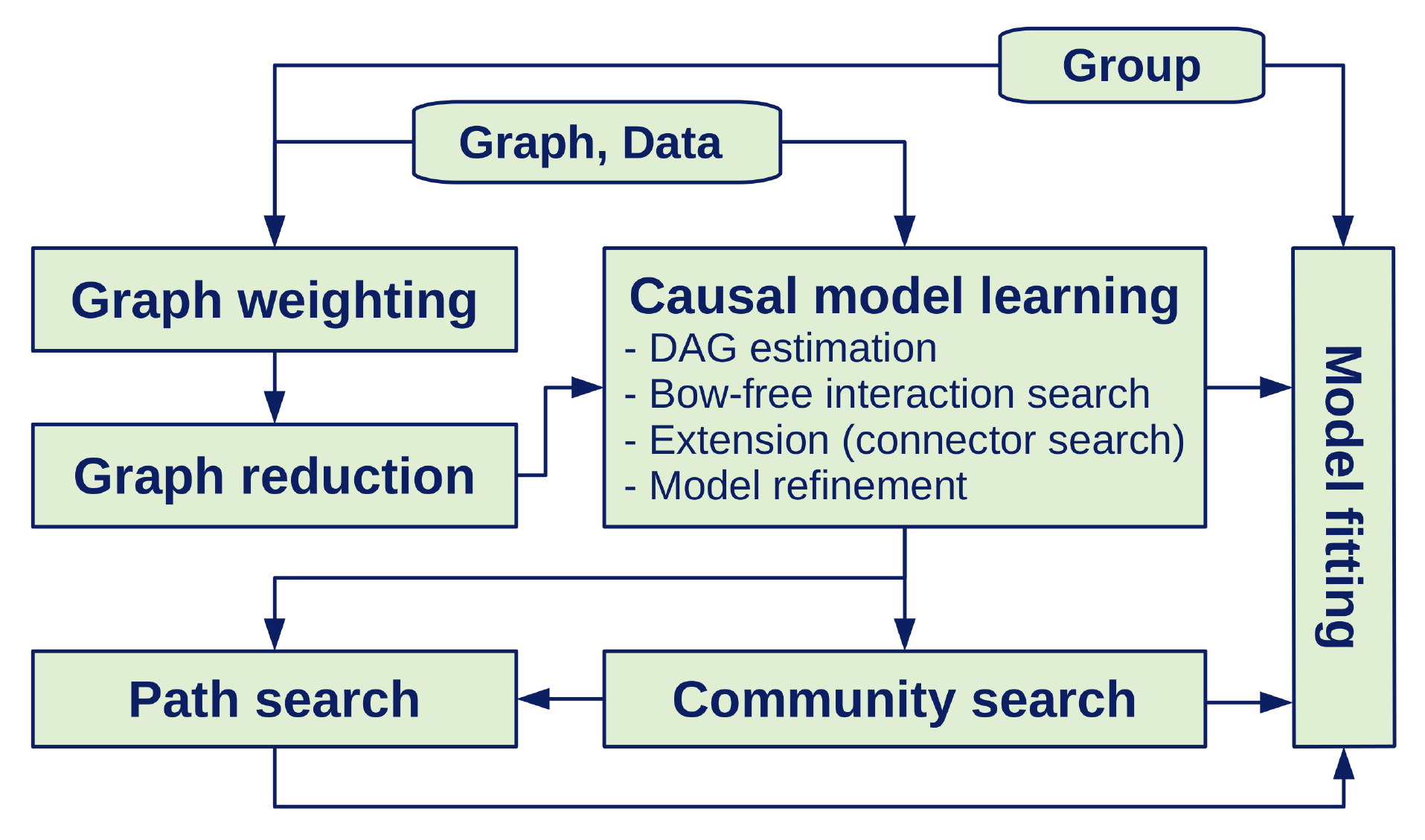}
\caption{\label{fig:workflow} \pkg{SEMgraph} basic analysis workflow.}
\end{figure}

\pkg{SEMgraph} uses \pkg{igraph} objects as input, although an 
internal SEM representation in \pkg{lavaan} syntax is also used by 
functions requiring model fitting. The user may manually change 
between these representations using simple conversion utilities. 
These functionalities reflect the four main steps of a typical \pkg{SEMgraph} 
workflow (see Figure~\ref{fig:workflow}), including: (i) data import and 
graph pre-processing; (ii) causal architecture learning; (iii) searching 
for (perturbed) network communities and paths; and (iv) model fitting. 
Beside the proposed scheme, the building blocks shown in 
Figure~\ref{fig:workflow} can be freely rearranged to generate custom 
workflows. A set of utilities for graph manipulation, format conversion, 
and visualization, complements the \pkg{SEMgraph} backbone, providing a 
self-sufficient toolkit for causal network analysis. The main goal of 
\pkg{SEMgraph} is the identification of critical players within the best 
causal model defined by three contextual sources of information that are 
simultaneously involved in model building and analysis: graph architecture, 
quantitative data, and the possible perturbing cause.\\
To achieve this goal, \pkg{SEMgraph} integrates different packages for 
model management and causal inference. Packages \pkg{igraph} \citep{igraph2006} 
and \pkg{lavaan} \citep{lavaan2012} provide the basic environment for 
model manipulation and fitting, while \pkg{glmnet} \citep{glmnet2012}, 
\pkg{dagitty} \citep{dagitty2016}, and \pkg{GGMncv} \citep{GGMncv2020} 
constitute the backbone for DAG estimation and BAP search. 
The employed methodologies are general enough to accept different graph 
types (e.g., directed, undirected, or mixed) and any kind of quantitative 
data, including bio-molecular, sequencing, and clinical data.

\subsection{Getting started with SEMgraph: SEM fitting functions} \label{sec:semfit}

\pkg{SEMgraph} comes with a collection of interactomes from commonly 
used biological databases, including KEGG \citep{KEGG2000}, STRING 
\citep{STRING2019}, and Reactome \citep{Reactome2020}.
Interactomes and data used in this work are available in the 
\pkg{SEMdata} data package at: 
\url{https://github.com/fernandoPalluzzi/SEMdata}.\\
Interactomes are stored as \lstinline{igraph} objects, so that they can be 
manipulated in \proglang{R} as any other graph. KEGG and Reactome are also 
present as a list of \lstinline{igraph} objects (\lstinline{kegg.pathways} 
and \lstinline{reactome.pathways}, respectively), each being a single 
pathway. In this section, we use KEGG pathways to build a SEM from an 
available graph (although the input can be any \lstinline{igraph} network 
object). As a first example, we could load a single pathway using:

\begin{lstlisting}
R> #load libraries
R> library(SEMgraph)
R> library(SEMdata)

R> graph <- properties(kegg.pathways$"Amyotrophic lateral sclerosis (ALS)")[[1]]
\end{lstlisting}
\begin{lstlisting}[stringstyle=\color{DarkGreen},
                   keywordstyle=\color{DarkGreen},
                   commentstyle=\color{DarkGreen}]
@Frequency distribution of graph components

  n.nodes n.graphs
1       1       16
2       3        1
3      32        1

Percent of vertices in the giant component: 62.7 %

  is.simple      is.dag is.directed is.weighted 
       TRUE        TRUE        TRUE        TRUE 
which.mutual.FALSE 
                47 @
\end{lstlisting}
Function \lstinline{properties()} takes an \lstinline{igraph} object and 
shows basic information about graph components, topology, and the presence 
of edge weights. In the example above, the KEGG pathway \emph{Amyotrophic 
Lateral Sclerosis} (ALS) is imported and the largest connected component 
is assigned to the \lstinline{graph} object in \lstinline{igraph} format. 
ALS RNA-seq expression data \citep{ALS2015} is downloaded, pre-processed, 
and stored in the \lstinline{alsData$exprs} object as a matrix of 
160 subjects $\times$ 17695 genes (with 139 ALS cases and 21 healthy 
controls). This is a high-dimensional data matrix, with the number of 
variables sensibly exceeding the number of observations ($p >> n$).\\
The three basic \pkg{SEMgraph} arguments are \lstinline{graph}, \lstinline{data}, 
and \lstinline{group}. Regarding quantitative data, we always suggest to apply 
some kind of correction method to relax the normality assumption required 
by SEM. While $log2$ or $ln$ transform are frequently used for count 
data (e.g., sequencing), we generally suggest the \emph{nonparanormal} 
transform implemented in the \lstinline{huge.npn()} function of the 
\proglang{R} package \pkg{huge} \citep{huge2012}.
\begin{lstlisting}
R> # ALS sample data
R> dim(alsData$exprs)       # ALS RNA-seq expression data
R> alsData$graph            # ALS input graph
R> table(alsData$group)     # {case = 1, control = 0} vector

R> # Nonparanormal transform
R> library(huge)
R> data.npn <- huge.npn(alsData$exprs)
\end{lstlisting}
In \pkg{SEMgraph}, the basic function for model assessment is 
\lstinline{SEMrun()}:
\begin{lstlisting}
R> sem0 <- SEMrun(graph = alsData$graph, data = data.npn)
\end{lstlisting}
\begin{lstlisting}[stringstyle=\color{DarkGreen},
                   keywordstyle=\color{DarkGreen},
                   commentstyle=\color{DarkGreen}]
@NLMINB solver ended normally after 25 iterations 
deviance/df: 10.92479  srmr: 0.2858233 @
\end{lstlisting}
This function maps data onto the input graph (removing possible identifiers 
inconsistencies), converts the input graph into a SEM, and fits the model 
using \pkg{lavaan}. For high-dimensional data, the shrinkage covariance 
proposed by \cite{Shrinkage2005} is applied to estimate the sample 
covariance $S$, as implemented in the \lstinline{cor.shrink()} function of the 
\pkg{corpcor} \proglang{R} package \citep{corpcor2017}. Model fitting 
results and the output graph are saved inside the \lstinline{sem} object. 
If the \lstinline{group} argument is omitted, \lstinline{SEMrun()} will only generate 
estimates for direct effects, as specified by the input \lstinline{graph}. 
Object \lstinline{sem0$fit} is a fitted model of class \lstinline{lavaan}, from 
which we can simply extract direct effect estimations with \lstinline{summary} 
or \lstinline{parameterEstimates}, as follows:
\begin{lstlisting}
R> est <- parameterEstimates(sem0$fit)
R> head(est)
\end{lstlisting}
\begin{lstlisting}[stringstyle=\color{DarkGreen},
                   keywordstyle=\color{DarkGreen},
                   commentstyle=\color{DarkGreen}]
@     lhs op   rhs    est    se      z pvalue ci.lower ci.upper
1 z10452  ~ z6647  0.037 0.079  0.466  0.641   -0.118    0.192
2  z1432  ~ z5606  0.397 0.069  5.741  0.000    0.261    0.532
3  z1432  ~ z5608  0.578 0.069  8.361  0.000    0.442    0.713
4  z1616  ~ z7132  0.245 0.110  2.236  0.025    0.030    0.461
5  z1616  ~ z7133 -0.036 0.110 -0.324  0.746   -0.251    0.180
6  z4217  ~ z1616 -0.074 0.079 -0.943  0.346   -0.229    0.080@
\end{lstlisting}
For gene networks, we always recommend using Entrez gene IDs, 
to avoid possible special characters or naming ambiguities. 
If the argument \lstinline{group} is given, group influence is modeled as 
an exogenous variable acting on every node, perturbing their activity. 
\begin{lstlisting}
R> sem1 <- SEMrun(alsData$graph, data.npn, alsData$group)
\end{lstlisting}
\begin{lstlisting}[stringstyle=\color{DarkGreen},
                   keywordstyle=\color{DarkGreen},
                   commentstyle=\color{DarkGreen}]
@NLMINB solver ended normally after 23 iterations 
deviance/df: 11.02558  srmr: 0.2747457 
Brown's combined P-value of node activation: 0 
Brown's combined P-value of node inhibition: 0.01061126 @
\end{lstlisting}
Also in this case, direct node-node effects, as well as group effects on 
nodes, can be inspected using \lstinline{parameterEstimates()}: 
\begin{lstlisting}
R> est1 <- parameterEstimates(sem1$fit)
R> head(est1)
\end{lstlisting}
\begin{lstlisting}[stringstyle=\color{DarkGreen},
                   keywordstyle=\color{DarkGreen},
                   commentstyle=\color{DarkGreen}]
@     lhs op   rhs    est    se      z pvalue ci.lower ci.upper
1 z10452  ~ group -0.150 0.078 -1.913  0.056   -0.303    0.004
2  z1432  ~ group -0.042 0.073 -0.578  0.563   -0.186    0.101
3  z1616  ~ group  0.025 0.079  0.315  0.753   -0.131    0.181
4   z317  ~ group  0.218 0.077  2.832  0.005    0.067    0.370
5  z4217  ~ group  0.176 0.078  2.273  0.023    0.024    0.328
6  z4741  ~ group  0.343 0.076  4.530  0.000    0.194    0.491@
\end{lstlisting}
Significant perturbed nodes can be viewed calling \lstinline{gplot()} on the 
output graph, as shown below. The resulting plot is shown in 
Figure~\ref{fig:E1_ALS_groupEffect}.
\begin{lstlisting}
R> # Convert Entrez identifiers to gene symbols
R> library(org.Hs.eg.db)
R> V(sem1$graph)$label <- mapIds(org.Hs.eg.db, V(sem1$graph)$name,
+                                column = 'SYMBOL',
+                                keytype = 'ENTREZID')
R> # Graph plot
R> gplot(sem1$graph)
\end{lstlisting}
High dimensionality can be troublesome not only due to a reduced sample 
size. Network size (i.e., the number of its nodes, $|V|$) may dramatically 
increase the computational demand, mainly during model parameters estimation. 
For large graphs ($|V| > 100$), standard error (SE) computation will be 
disabled and parameter estimates will be computed through residual iterative 
conditional fitting (RICF), from the \proglang{R} package \pkg{ggm} 
\citep{ggm2020}. Group effect P-values are computed by randomization of 
group labels, using the \pkg{flip} \proglang{R} package \citep{flip2018}. 
The RICF mode is either automatically enabled when $|V| > 100$ (this limit 
can be changed using the \lstinline{limit} argument in \lstinline{SEMrun()}, to 
enforce standard SE estimation) or manually called using the \lstinline{algo} 
argument: 
\begin{lstlisting}
R> ricf1 <- SEMrun(alsData$graph, data.npn, alsData$group, algo = "ricf")
\end{lstlisting}
\begin{lstlisting}[stringstyle=\color{DarkGreen},
                   keywordstyle=\color{DarkGreen},
                   commentstyle=\color{DarkGreen}]
@RICF solver ended normally after 2 iterations 
deviance/df: 10.26773  srmr: 0.2747457 
Brown's combined P-value of node activation: 0 
Brown's combined P-value of node inhibition: 0.007574838 @
\end{lstlisting}
As for the basic (i.e., lavaan-based) algorithm, the command 
\lstinline{gplot(ricf1$graph)} can be used with the \lstinline{gplot()} function 
to plot node perturbation. 
The RICF is an efficient iterative algorithm that can be implemented through 
least squares, with the advantage of clear convergence properties, yielding 
exact MLE after the first iteration whenever the MLE is available in closed 
form \citep{RICF2009}.\\
Both lavaan-based and RICF-based fitting show two important results. 
Firstly, the randomization approach leads to a perturbation 
estimation that is not significantly different from the asymptotic one 
(model fitting and overall perturbation is left unaltered by both RICF 
and the randomization procedure). Secondly, both functions detect 
significant network perturbation (mainly activation), but no acceptable 
fitting (see Section~\ref{sec:semcausal} for model refinement).\\
In addition to node perturbation, \pkg{SEMgraph} enables edge perturbation 
estimation via the two-groups SEM implemented in \lstinline{SEMrun}, setting 
the \lstinline{fit} argument to 2 groups (see Section~\ref{sec:sem2gr} for 
details):
\begin{lstlisting}
R> sem2 <- SEMrun(alsData$graph, data.npn, alsData$group, fit = 2)
\end{lstlisting}
\begin{lstlisting}[stringstyle=\color{DarkGreen},
                   keywordstyle=\color{DarkGreen},
                   commentstyle=\color{DarkGreen}]
@Estimating optimal shrinkage intensity lambda (correlation matrix): 0.4313 
NLMINB solver ended normally after 30 iterations 
deviance/df: 5.295486  srmr: 0.2785664 
Brown's combined P-value of edge activation: 0.001049916 
Brown's combined P-value of edge inhibition: 0.9570024 @
\end{lstlisting}
\begin{figure}[t!]
\centering
\includegraphics[scale=0.45]{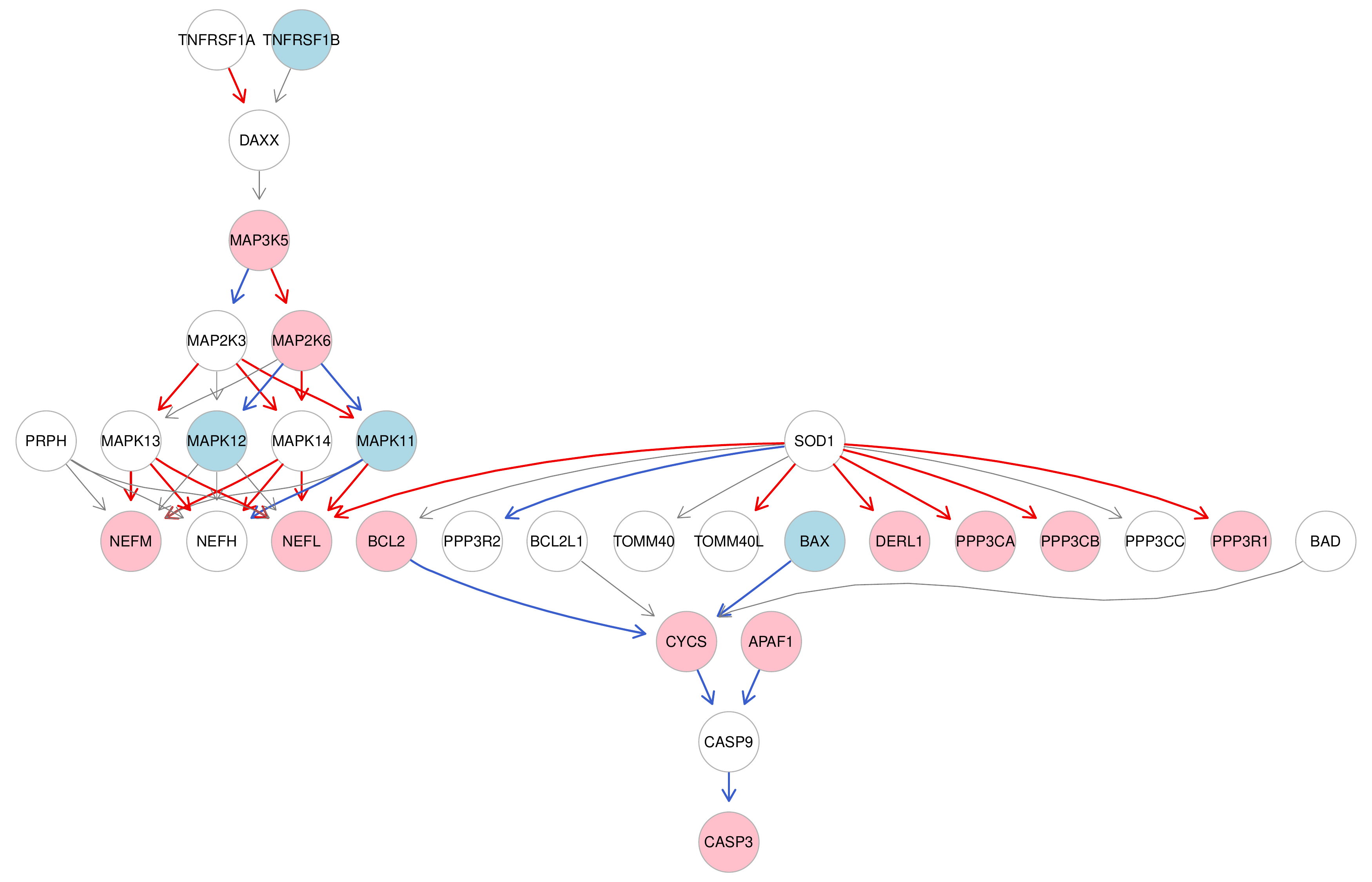}
\caption{\label{fig:E1_ALS_groupEffect} 
Estimated group effects on nodes and direct effects. The graph shows 
differentially regulated nodes (DRNs) as ALS-activated (pink-shaded) or 
ALS-inhibited (blue-shaded) variables. White nodes do not show significant 
variation in ALS, respect to healthy controls. Significant direct effects 
are shown in either red (activated) or blue (inhibited), while gray direct 
common effects are not significant.}
\end{figure}
In accordance with node perturbation, we observe a predominant global edge
activation. As for node-level testing, edge perturbation can be plotted 
through the command \lstinline{gplot(sem2$graph)}. The list of DRNs and DREs 
can be extracted from the objects \lstinline{sem1$gest} and \lstinline{sem2$dest}, 
respectively:
\begin{lstlisting}
R> DRN <- sem1$gest[sem1$gest$pvalue < 0.05,]
R> nrow(DRN); head(DRN)
\end{lstlisting}
\begin{lstlisting}[stringstyle=\color{DarkGreen},
                   keywordstyle=\color{DarkGreen},
                   commentstyle=\color{DarkGreen}]
@[1] 16
> head(DRN)
      lhs op   rhs   est    se     z pvalue ci.lower ci.upper
4     317  ~ group 0.218 0.077 2.832  0.005    0.067    0.370
5    4217  ~ group 0.176 0.078 2.273  0.023    0.024    0.328
6    4741  ~ group 0.343 0.076 4.530  0.000    0.194    0.491
8    4747  ~ group 0.223 0.062 3.611  0.000    0.102    0.344
9   54205  ~ group 0.188 0.067 2.789  0.005    0.056    0.319
10   5530  ~ group 0.160 0.072 2.224  0.026    0.019    0.301@
\end{lstlisting}
\begin{lstlisting}
R> DRE <- sem2$dest[sem2$dest$pvalue < 0.05,]
R> nrow(DRE); head(DRE)
\end{lstlisting}
\begin{lstlisting}[stringstyle=\color{DarkGreen},
                   keywordstyle=\color{DarkGreen},
                   commentstyle=\color{DarkGreen}]
@[1] 3
> head(DRE)
     lhs op   rhs d_est  d_se   d_z pvalue d_lower d_upper
28  5532  ~  6647 0.449 0.227 1.983  0.047   0.005   0.893
30  5534  ~  6647 0.584 0.229 2.547  0.011   0.135   1.034
34  5603  ~  5606 0.496 0.239 2.073  0.038   0.027   0.965@
\end{lstlisting}
The current model yields 16 DRNs and 3 DREs. With increasing $|V|$, 
also the edge perturbation estimation could be computationally intensive. 
For large graphs (by default, $|V| > 100$), edge perturbation is estimated 
using a constrained gaussian graphical model (GGM) and de-sparsified 
P-values, as implemented in the \pkg{GGMncv} package \citep{GGMncv2020}. 
Also in this case, the canonical (i.e., lavaan-based) perturbation 
estimation can be enforced by increasing the \lstinline{limit} argument.

\subsection{Total effect estimation} \label{sec:semace}

As anticipated in Section~\ref{sec:effdec}, total effect (TE) estimation 
could be a key tool to search for perturbed routes conveying information 
inside a complex network. Biological signaling pathways provide a paradigmatic 
example of this propagation inside the cell regulatory network. A ligand 
interacts with a cell surface receptor (\emph{source}), starting the 
information flow that is propagated and modulated by second messengers, 
enzymes and chaperones (\emph{connectors}) through the cytoplasm to the 
cell nucelous, where specific factors (\emph{sinks}) are either activated 
or inhibited, regulating transcription, replication, cell development, 
and fate. This directional information flow can be computationally 
represented by a DAG, where the TE can be evaluated with a single 
comprehensive estimation as an average causal effect (ACE). Function 
\lstinline{SEMace()} converts the input graph into a DAG and computes 
ACEs between every possible source-sink node pair, using the optimal 
adjustement set (O-set) procedure described in Section~\ref{sec:effdec}:
\begin{lstlisting}
R> ace <- SEMace(graph = alsData$graph, data = data.npn, method = "BH")
R> ace <- ace[order(abs(ace$z), decreasing = TRUE),]
R> nrow(ace); head(ace)
\end{lstlisting}
\begin{lstlisting}[stringstyle=\color{DarkGreen},
                   keywordstyle=\color{DarkGreen},
                   commentstyle=\color{DarkGreen}]
@[1] 11
    sink op source    est    se      z pvalue ci.lower ci.upper
4   4747 <-   6647  0.514 0.063  8.113      0    0.390    0.639
14   836 <-    317  0.472 0.061  7.737      0    0.352    0.592
5  79139 <-   6647  0.522 0.068  7.723      0    0.390    0.655
7   5532 <-   6647  0.521 0.068  7.700      0    0.389    0.654
10  5535 <-   6647 -0.462 0.070 -6.565      0   -0.600   -0.324
3    836 <-   6647  0.430 0.067  6.433      0    0.299    0.561@
\end{lstlisting}
In this example, there are 11 significant ACEs, ordered by decreasing 
$z$ scores. Function \lstinline{SEMpath()} allow us to evaluate any of
them as an independent model. The following code shows fitting and
node perturbation estimation for the sixth directed path in the example
above, connecting SOD1 (Entrez ID: 6647) and CASP3 (Entrez ID: 836):
\begin{lstlisting}
R> source <- as.character(ace$source[6])
R> sink <- as.character(ace$sink[6])
R> path <- SEMpath(alsData$graph, data.npn, alsData$group,
+                  from = source, to = sink, 
+                  path = "directed",
+                  verbose = TRUE)
\end{lstlisting}
\begin{lstlisting}[stringstyle=\color{DarkGreen},
                   keywordstyle=\color{DarkGreen},
                   commentstyle=\color{DarkGreen}]
@NLMINB solver ended normally after 12 iterations 
deviance/df: 24.52598  srmr: 0.2067487 
Brown's combined P-value of node activation: 3.724367e-06 
Brown's combined P-value of node inhibition: 0.9286749 @
\end{lstlisting}
Argument \lstinline{path = "directed"} considers every directed path connecting 
the source-sink pair. This argument can be also set to \lstinline{"shortest"}, 
to consider shortest paths only. Argument \lstinline{verbose = TRUE} shows 
the position of the selected path within the input network. 
Function \lstinline{pathFinder()} can be used to extract all the directed paths 
whose source-sink pairs share a significant ACE and evaluate each of them 
as an independent SEM:
\begin{lstlisting}
R> paths <- pathFinder(alsData$graph, data.npn, alsData$group, ace = ace)
\end{lstlisting}
Argument \lstinline{ace} allows the user to specify an existing data.frame of 
ACEs, while \lstinline{group} can be skipped if one is just interested in path 
fitting (i.e., no node perturbation test is performed).

\subsection{Gene set analysis} \label{sec:semgsa}

When the perturbation of a biological network is associated to a disease, 
a systematic review of known biological networks may give important clues 
about the functional implication and molecular mechanisms of disease 
associated alterations. To this end, \pkg{SEMgraph} provides tools for 
gene set analysis (GSA), enabling fast and accurate testing at gene and 
pathway level. The core of SEM-based GSA methodology is implemented in 
the RICF-based method implemented in \lstinline{SEMrun()}. In addition to 
node-level and model fitting estimates, \lstinline{SEMrun()} RICF-based 
algorithm computes three global measures of pathway perturbation:
\begin{itemize}
\item \emph{Total pathway perturbation} adjusted by model covariances ($D$). 
$D$ is the sum of residual decorrelated mean differences between groups 
and its sign determines pathway activation or inhibition. It is calculated 
as the square root of the Mahalanobis distance \citep{mahalanobis1936} 
of group mean vector $D^2 = (\bar{y}_1 - \bar{y}_0)^{T} \, S^{-1} \, 
(\bar{y}_1 - \bar{y}_0) / p$, replacing the observed precision matrix 
$S^{-1}$ with the estimated SEM precision matrix $\hat\Sigma^{-1} = 
(I - \hat{B})^{T} \, \hat\Psi^{-1} \, (I - \hat{B})$.
\item \emph{Total perturbation accumulated} by sink nodes ($A$). The 
perturbation accumulation of the $j$-th target gene is given by its group 
mean difference weighted by the sum of incoming effects $\beta_{j+}$ of 
its upstream (i.e., ancestor) genes. Thus $A$ corresponds to the linear 
combination of the incoming effects on every pathway sink and its sign 
determines overall perturbation accumulation in terms of activation or 
inhibition.
\item \emph{Total perturbation emitted} by source nodes ($E$). Similarly 
to $A$, $E$ is calculated as the linear combination of the outgoing effects 
$\beta_{+k}$ of every ancestor gene on downstream (i.e., descendant) genes, 
using the sum of outgoing effects as weights. The sign of $E$ determines 
the overall perturbation emission in terms of activation or inhibition.
\end{itemize}
These three measures are formally defined as follows:
\begin{equation} \label{eq:ricfd}
D = (\bar{y}_1 - \bar{y}_0)^{T} \, \hat\Sigma^{-1/2} \, \nu = 
\frac{\sum_{j}^{} (\bar{z}_{j1} - \bar{z}_{j0})}{\sqrt{p}}
\end{equation}
\begin{equation} \label{eq:ricfd}
A = (\bar{y}_1 - \bar{y}_0)^{T} \, \hat{B} \, \nu = 
\frac{\sum_{j}^{} \hat\beta_{j+} \, (\bar{y}_{j1} - \bar{y}_{j0})}{\sqrt{p}}
\end{equation}
\begin{equation} \label{eq:ricfd}
E = (\bar{y}_1 - \bar{y}_0)^{T} \, \hat{B}^T \, \nu = 
\frac{\sum_{k}^{} \hat\beta_{+k} \, (\bar{y}_{k1} - \bar{y}_{k0})}{\sqrt{p}}
\end{equation}
where $z = \hat\Sigma^{-1/2} \, y$ represents the decorrelated data 
$y$, $\hat{B}$ is the matrix of the estimated beta coefficients, and 
$\nu^T = (1, 1, \dots, 1)/\sqrt{p}$.\\
While $A$ and $E$ are suited for describing directed (hierarchical) networks, 
such as signaling pathways, $D$ can describe perturbation in both directed 
and undirected networks. Permuted P-values of the aggregated statistics 
$T = (D, A, E)$ for directed graphs, or $T = D$ for undirected graphs, 
are evaluated by comparing the observed values of $T$ with their random 
resampling distribution after a sufficiently high number of case/control 
labels permutations. In \pkg{SEMgraph}, this is implemented using the 
\proglang{R} package \pkg{flip} \citep{flip2018}. For large networks 
($p >> n$), accurate P-value estimations are possible with no need for 
a large number of permutations (\lstinline{SEMrun()} makes 5000 permutations), 
using the moment based approximation proposed by \cite{Permutation2015}. 
Once the empirical distribution of the permuted statistic $T$ is obtained, 
the two-sided P-values are computed from the normal distribution with 
mean and standard deviation estimated by the empirical distribution. 
These estimates can be viewed at the top three lines of the \lstinline{gest} 
object:
\begin{lstlisting}
R> ricf <- SEMrun(alsData$graph, data.npn, alsData$group, algo = "ricf")
R> head(ricf$gest)
\end{lstlisting}
\begin{lstlisting}[stringstyle=\color{DarkGreen},
                   keywordstyle=\color{DarkGreen},
                   commentstyle=\color{DarkGreen}]
@     Test    Stat tail      pvalue
D       t  2.4439   >< 0.015854612
A       t  3.3784   >< 0.000749345
E       t -1.0422   >< 0.298113627
317     t  2.8143   >< 0.006438137
572     t -1.3373   >< 0.199006357
581     t -2.0472   >< 0.043156472@
\end{lstlisting}
In this case, the ALS sinks accumulate a significant perturbation, causing 
their activation (P-value($A$) < 0.05 and statistic > 0), as well as a 
global network activation (P-value($D$) < 0.05 and statistic > 0). 
Conversely, source perturbation emissions are not significant 
(P-value($E$) > 0.05). From Figure~\ref{fig:E1_ALS_groupEffect}, 
it looks evident how the majority of sinks are up-regulated. However, 
for larger and more complex networks, global perturbation significance 
and direction could be harder to spot by eye. Notably, although sources 
do not show significant alterations, perturbation is accumulated through 
the routes traversing connectors, to the sinks, activating them in the 
ALS respect to healthy subjects. Sink perturbation can be used as a measure 
of the alteration specificity. Sources are often receptors or messengers 
involved in more biological processes. On the other hand, sinks are effectors 
specific for a restricted set of functions, hence directly connected to 
the functional alterations characterizing the diseased phenotype. 
Function \lstinline{SEMgsa()} uses the RICF method to iteratively apply the 
GSA on a list of gene networks (in our example, KEGG signaling pathways):
\begin{lstlisting}
R> n <- unlist(lapply(1:length(kegg.pathways),
+              function(x) vcount(kegg.pathways[[x]])))
R> blacklist <- which(n < 5 | n > 500)
R> length(blacklist)
R> pathways <- kegg.pathways[-blacklist]
R> GSA <- SEMgsa(pathways, data.npn, alsData$group, method = "BH", alpha = 0.05)
\end{lstlisting}
Every pathway is listed in the \lstinline{GSA$gsa} data.frame, reporting 
size, DRN number, P-values for $D$, $A$, and $E$ (i.e., \lstinline{pD}, \lstinline{pA}, 
and \lstinline{pE}, respectively), and the Fisher\textquotesingle s combination 
of them (\lstinline{p.value}). In addition, the list \lstinline{GSA$DRN} contains 
a vector of DRN IDs for each pathway, selected with P-value < \lstinline{alpha} 
after Benjamini-Hochberg correction (\lstinline{method = "BH"}). In this example, 
we used the \lstinline{kegg.pathways} list, though any list of \lstinline{igraph} 
network objects can be passed.

\section{Causal structure learning} \label{sec:semcausal}

In biological systems, curated networks rarely provide a complete 
explanation of data variability, often leading to a poor SEM fitting. 
This is exactly what happened when we fitted RNA-seq ALS data onto the 
ALS pathway provided by KEGG. In this case, the known ALS model is able 
to detect significantly perturbed nodes and edges, but a significant 
proportion of data variability is still unexplained, as shown by the 
global fitting statistics (deviance/df and SRMR). \pkg{SEMgraph} main goal 
is to learn the causal structure from data, applying the best tradeoff 
between model fitting and perturbation.\\
Generally, causal inference applied to complex biological systems relay 
on models that are either a priori conceptual constructs given by the expert 
or curated knowledge-based networks from biological repositories (typically 
molecular, genetic, or protein-protein interaction databases) 
\citep{Network2020, NetworkMed2011}. 
On the other hand, fully data-driven networks provide exploratory structures 
unravelling hidden knowledge, although they can be deeply affected by 
technical variability, and the specific method used to build them often 
results in very different or irreproducible networks \citep{Network2020}. 
\pkg{SEMgraph} offers three methods to cope with these limitations, improving 
the initial model by leveraging on both knowledge-based and data-driven 
procedures. Firstly, \lstinline{SEMdag()} uses data and topological information 
from the input network to estimate the optimal directed (i.e., causal) 
edge backbone. In addition, \lstinline{SEMbap()} uses missing edges from the 
input graph to search for bidirected edges (i.e., covariances)  based on
 conditional independence tests, removing possible latent sources of 
confounding, encoded in the estimated covariance matrix. Finally, 
\lstinline{extendGraph()} uses external interactomes (e.g., from a chosen 
biological database) and observed data to extend the input graph with 
new connectors. The next sections will dive into the details of these 
core functions.

\subsection{DAG estimation} \label{sec:semdag}

\lstinline{SEMdag()} estimates the causal structure of a DAG, inferring the 
parent set of each variable, given data. However, the causal DAG is 
generally not identifiable, while only its Markov equivalence class is 
(i.e., the list of all equivalent DAGs). Recent work established that 
exact identification, and not just an equivalent class, is possible 
under specific assumptions, including nonlinearity with additive errors, 
linearity with non-Gaussian errors, and linearity with errors of equal 
variance \citep{graphLearn2017, Causality2018}. A key observation, 
under the error equal variance assumption, is that ordering among 
conditional variances implies data-driven identifiability. After 
estimating the (top-down or bottom-up) ordering of a graph, its unique 
causal structure can then be inferred \citep{EqVarDAG2019}. Alternatively, 
the natural ordering of a biological network (e.g., a gene or protein 
interaction network) could be typically obtained from \emph{a priori} 
information (e.g., from signaling pathway or transcription factor binding 
databases) \citep{KEGG2000, Reactome2020} or inferred using expression 
quantitative trait loci in the neighborhood of transcription start sites 
of known genes (cis-eQTL), used as causal anchors \citep{eQTL2019}.\\
The problem of estimating the skeleton of a DAG can be seen in terms of 
penalized likelihood, as suggested by \cite{DAG2010}. Assuming that the 
topological ordering of the variables (nodes) $Y_1 < Y_2 < \dots < Y_p$ 
is known, where the relation $k < j$ is interpreted as \textquotedbl node 
$k$ precedes node $j$\textquotedbl (i.e., there is an acyclic path from 
node $k$ to node $j$). Then, the estimate of the graph adjacency matrix 
$A$ can be solved by $p - 1$ LASSO (Least Absolute Shrinkage and Selection 
Operator) regressions of the $j$-th outcome variable on the predictor 
variables $k = 1 \mathrm{,} \, \dots \mathrm{,} \, (j - 1)$ in the order 
list:
\begin{equation} \label{eq:l1lasso}
\hat{A}_{j; \, 1:j-1} = \argmin_{\beta \, \in \, \mathbb{R}^{j - 1}} \, 
\left\{\frac{1}{n} \sum_{i = 1}^{n} (y_{ij} - y_{i}^{T} \beta)^2 + 
\lambda_j \sum_{k = 1}^{j - 1} w_{jk} |\beta_{jk}| \right\} 
\qquad (j = 2, \dots, p)
\end{equation}
where $\hat{A}_{j; \, 1:j-1}$ denotes the first 1 to $(j-1)$ elements of 
the $j$-th column of $A$, and $\lambda_j$ is the tuning parameter for each 
LASSO regression problem. Separate penalty factors $w_{jk}$ can be applied 
to each coefficient to allow differential shrinkage. If $w_{jk} = 0$ for 
some variables, it implies no shrinkage and those variables are always 
included in the selected model.\\
Function \lstinline{SEMdag()} converts the input graph in a DAG, sorts its 
nodes in a topological order, and solves the $(j = 2, \dots, p)$ LASSO 
problems, using the extremely fast cyclic coordinate descent optimization 
algorithm, implemented in the \proglang{R} package \pkg{glmnet} 
\citep{Regularization2010}. Using penality weights 0 (i.e., edge present) 
and 1 (i.e., missing edge) for the DAG adiacency matrix ensures that 
input DAG edges will be retained in the final model. Function 
\lstinline{SEMdag()} takes an input \lstinline{graph}, a \lstinline{data} 
matrix or data.frame, and a reference directed interactome, if available: 
\begin{lstlisting}
R> DAG <- SEMdag(graph = alsData$graph, data = data.npn, gnet = kegg,
+                d = 2, beta = 0, lambdas = NA, verbose = FALSE)
\end{lstlisting}
Argument \lstinline{gnet} is used to specify the reference interactome as an 
\lstinline{igraph} object. The reference network should ideally encompass 
the current knowkedge domain, providing the largest possible framework 
in which the input model is embedded. In our ALS example, we used 
\lstinline{kegg} as a reference. This means that every added directed 
interaction is checked in KEGG. If a reference is not available, 
the \lstinline{gnet} argument can be skipped and the DAG estimation will be 
fully data-driven (no reference-based validation is required). If 
\lstinline{gnet} is not \lstinline{NULL}, argument \lstinline{d} determines the maximum 
geodesic distance between two nodes in the interactome, to consider the 
inferred interaction between the same two nodes in the DAG as validated. 
For instance, if \lstinline{d = 2}, two interacting nodes in the output DAG 
must either share a direct interaction or being connected through at most 
one mediator in the reference interactome (in general, at most \lstinline{d - 1} 
mediators are allowed). Typical \lstinline{d} values include \lstinline{2} (at most 
one mediator), \lstinline{mean_distance(gnet)}, or \lstinline{mean_distance(graph)} 
(i.e., the average shortest path length for the reference network and the 
input graph, respectively). Argument \lstinline{beta} (by default, 
\lstinline{beta = 0}) is the threshold LASSO coefficient which retains only 
those variables for which the absolute value of the LASSO coefficients 
exceed the threshold (i.e., higher beta values correspond to sparser 
output DAGs) \citep{ThresholdLASSO2009}. 
Argument \lstinline{lambdas} can be used to specify a vector of LASSO $\lambda$ 
values. As an alternative, cross-validation ($|V| >$ 100) or BIC-based 
($|V| \leq$ 100) optimal lambdas for each response variable will be selected. 
If \lstinline{lambdas} is \lstinline{NULL}, the \pkg{glmnet} default is used, while 
if \lstinline{lambdas} is \lstinline{NA} (default), a tuning-free scheme is enabled 
by fixing \lstinline{lambdas = sqrt(log(p)/n)}, as suggested by \cite{invcov2015}. 
Finally, enabling \lstinline{verbose}, the output DAG (object \lstinline{DAG$dag}) 
will be plotted: blue edges are the ones imported from the input graph, 
and red edges are the interactions inferred from data. 

\subsection{BAP deconfounding} \label{sec:sembap}

Function \lstinline{SEMbap()} provides local DAG fit evaluation and data 
de-correlation methods through BAP exhaustive search from an input DAG. 
The idea behind this approach is based on the causal interpretation of 
BAPs. Two connection types characterize a BAP: \emph{directed} (direct 
effect) and \emph{bidirected} (covariance). 
A directed edge from node $Y_j$ to node $Y_k$ represents a direct causal 
effect of $Y_j$ on $Y_k$. A bidirected edge between $Y_j$ and $Y_k$ can 
be interpreted as a latent variable (LV) acting on both $Y_j$ and 
$Y_k$. This LV may be the cause of a correlation between observed variables; 
i.e., the LV is an unobserved confounder \citep{Causality2000}. This 
correlation can be misleading and can only be correctly explained if the 
presence of the LV that produce the confounding effect is evaluated.\\
We can use Shipley\textquotesingle s independent d-separation local tests
(see Section~\ref{sec:semfit}) for DAG evaluation. As stated by
\cite{SEM2000} (p. 217): \textquotedbl Because the individual tests implied 
by the basis set $B_U$ are mutually independent, each one can be tested 
separately at a significance level of $\alpha/B$, where $B$ is the number 
of tests performed, following a Bonferroni test logic. In this way, 
lack-of-fit in the whole model can be decomposed into lack-of-fit involving 
pairs of variables\textquotedbl. 
Extensions to DAGs with correlated errors (i.e., BAPs) can also be obtained. 
There is currently no method to obtain a mutually independent basis 
set $B_U$ for a BAP. However, each pair of nonadjacent variables in a 
BAP model implies that there is some set of other observed variables 
that, on conditioning, will make the two nonadjacent variables independent. 
Hence, it is always possible to obtain a minimal set $B_M = \{ Y_j 
\perp Y_k \, | \, \mathrm{min}(S) \}$ consisting of each nonadjacent pair 
$(Y_j; Y_k)$ in the model, and the smallest conditioning set $S$ that makes 
these two variables independent and a significance level $\alpha$ 
\citep{BAP2002}. \\
Significant local tests do not indicate a specific direction of causality, but
provide information about which part of a DAG is not supported by the observed
data, identifying the local misspecification given by the structural assumptions
implied by the DAG, that may substantially alter the observed data variability.
We assume that the model misspecification is determined by unobserved 
confounders (i.e., LVs). These LVs may include, for example,
biomarkers that are not  observed in experimental chips, environmental
variables, or underlying populations among experimental samples.\\
In summary, BAP search could be performed with d-separation or conditional
independence (CI) tests between all pairs of variables with missing connection
in the input DAG. A BAP is then built by adding a bidirected edge (i.e., bow-free 
covariance) to the DAG when there is an association between them
at a significance level $\alpha$, after multiple testing correction.\\
Intuitively, it would be impossible to evaluate a causal DAG if the nuisance
LVs, encoded in the bow-free covariances, are not properly removed.
If the BAP represents a good compromise between map accurateness 
and non-identified factors, and the implied population precision matrix 
$\Psi^{-1}$ is know, we can adjust (or de-correlate) the
observed variables $Y$ via, in the matrix form of equations (\ref{eq:semsys})
of Section~\ref{sec:sembasics}:
\begin{equation} \label{eq:adjLogl}
\Psi^{-1/2}Y=\Psi^{-1/2}(B\,Y+U) \,\,\,\, \text{s.t.}\,\,\,\,Z = A\,Z+ D
\end{equation}
where $A=\Psi^{-1/2}B\Psi^{1/2}$, $Z=\Psi^{-1/2}Y$ and $D=\Psi^{-1/2}U$.
By definition this model assumes independence among error terms, i.e., 
is a DAG: $\mathrm{cov}(D) = D^{T}D = I$, and considering that
$\mathrm{det} (\Sigma) = \mathrm{det} (D) = 1$, the
log-likelihood function (see \ref{eq:logl}) in Section~\ref{sec:sembasics})
is reduced to: $ -\frac{1}{2} \mathrm{tr}[(Z - A \, Z) \, (Z - A \, Z)^T]$.

The population precision matrix $\Psi^{-1}$ is not known, therefore the adjusted
(de-correlate) variables $Z = \Psi^{-1/2}Y$ should be estimated from data.
We suggest a two-step procedure: (i) fitting the constrained precision matrix 
$\Psi^{-1}$ with null (zero) pattern corresponding to the DAG edges and the 
null ($P > \alpha$) edges after the local d-separation or CI screening, and 
(ii) removing by conditioning out from the observed data the latent triggers 
responsible for the nuisance edges by the spectral decomposition of the fitted 
precision matrix $\hat\Psi^{-1} = VLV^{T}$, from which we get the adjusted
(de-correlate) matrix, $Z = VL^{\frac{1}{2}} V^{T}Y$. \\
Using $Z$  as new data, may lead to an improvement of DAG fitting, encoded
in the matrix $B$. Since the confounding correlation in Z vanishes, we find that
this de-correlation step is able to substantially decrease DAG badness of fit
indices, such as C or SRMR, applying the best tradeoff between global model
fitting and local statistical significance of path coefficients \textbf{[reference]}.\\
In \lstinline{SEMgraph}, the constrained estimation of the precision matrix 
$\Psi^{-1}$ and the spectral decomposition are implemented using the 
\lstinline{constrained()} function of the \proglang{R} package \lstinline{GGMncv} 
\citep{GGMncv2020} and the \lstinline{eigen()} \proglang{R} function 
\citep{R2020}, respectively. The \lstinline{SEMbap()} function has the following syntax:
\begin{lstlisting}
R> BAP <- SEMbap(graph = alsData$graph, data = data.npn, 
+                method = "bonferroni",  alpha = 0.05,
+                limit = 30000, verbose = FALSE)
\end{lstlisting}
Argument \lstinline{alpha} determines the significance level after d-separation 
testing (by default, \lstinline{alpha = 0.05}). Argument \lstinline{limit} 
corresponds to the number of missing edges beyond which multithreading 
is enabled to reduce the computational burden. 
Finally, the \lstinline{verbose = TRUE} option plots the intermediate covariance 
and LV structure used for the BAP search. The output of \lstinline{SEMbap()} 
are four objects: the BAP (i.e., the union between the input graph 
and the bow-free covariance graph), the covariance graph, the directed 
graph of LVs underlying significant covariances (i.e., the canonical graph, 
where bidirected $Y_j \leftrightarrow Y_k$ edges are substituted by directed 
edges $Y_j \leftarrow \mathrm{LV} \rightarrow Y_k$), and a data.frame of 
the adjusted (i.e., de-correlated) data matrix $Z$.

\subsection{Graph extension} \label{sec:semext}

Both directed (causal) edges inferred by \lstinline{SEMdag()} and covariances 
(i.e., bidirected edges) added by \lstinline{SEMbap()}, highlight emergent 
hidden topological proprieties, absent in the input graph. Estimated 
directed edges between nodes $Y_j$ and $Y_k$ are interpreted as either 
direct links or direct paths mediated by connector nodes. Covariances 
between any two bow-free nodes $Y_j$ and $Y_k$ may hide causal relationships, 
not explicitly represented in the current model. If this latent cause 
exists, the presence of a covariance can be considered as a potential 
source of model misspecification, and can be either data-driven adjusted 
or recovered from a reference database. Missing information could be 
recovered from a large interaction database, revealing two main types of 
system elements not explicitly represented by the current model: hidden 
mediators within a directed path, and hidden variables (e.g., LVs) masked 
by a covariance. Function \lstinline{extendGraph()} leverage on these concepts 
to extend a causal model, importing new directed edges and connectors 
(i.e., mediators) from a given reference network:
\begin{lstlisting}
R> ext <- extendGraph(g = list(DAG$dag, DAG$dag.red), data = data.npn,
+                     gnet = kegg, verbose = FALSE)
\end{lstlisting}
This function takes three input graphs: the first is the input causal model 
(i.e., a directed graph), the second can be either a directed or undirected 
graph, providing a set of connections to be checked against the reference 
network (i.e., the third input). In the example above, we used the DAG 
estimated by \lstinline{SEMdag()} (object \lstinline{DAG$dag}) and the new estimated 
edges (object \lstinline{DAG$dag.red}) as first and second input, respectively. 
The reference network (\lstinline{gnet = kegg} in our example) should have 
weighted edges, corresponding to their interaction P-values, as an edge 
attribute \lstinline{E(kegg)$pv} (see Section~\ref{sec:semWeightRed}).
Then, connections in the second graph will be substituted by known 
connections from the reference network, intercepted by the minimum-weighted 
shortest path found among the equivalent ones by the 
Dijkstra\textquotesingle s algorithm, as implemented in the \pkg{igraph} 
function \lstinline{all_shortest_paths()}. If the reference netwok has unweighted 
edges, one random shortest path will be chosen among the equivalent ones.
The interactions imported from the reference network will be added to the 
first causal graph. If the reference is an undirected network,  an extended
undirected graph will be inferred.
The resulting graph is saved in the \lstinline{ext$Ug} object. The whole process 
may lead to the discovery of new paths of information flow, from network 
sources to sinks, and the presence of novel connectors between them. 
Since added nodes can already be present in the input graph, network 
extension may create cross-connections between old and new paths and their 
possible closure into circuits.

\subsection{Model estimation strategies} \label{sec:semStrategies}

\begin{figure}[t!]
\centering
\includegraphics[scale=0.38]{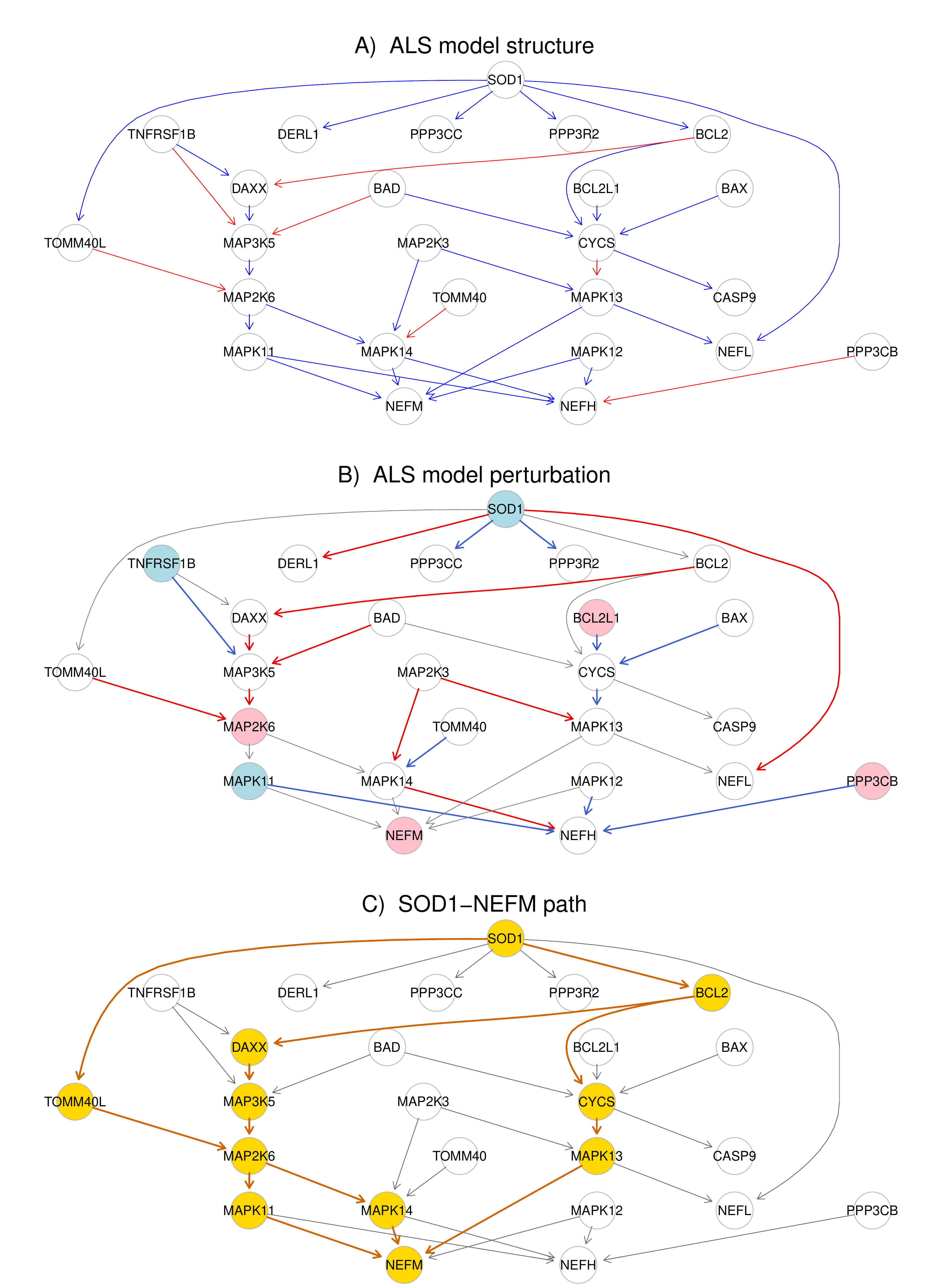}
\caption{\label{fig:improvedModel} ALS improved model. \textbf{Panel A} 
shows the output model structure, as generated by \lstinline{modelSearch()}. 
Added edges are highlighted in red, while blue edges are maintained from 
the input ALS graph. \textbf{Panel B} shows node-level perturbation, 
estimated by \lstinline{SEMrun()}: pink nodes are activared, while lightblue 
nodes are inhibited. Edges are coloured according to their significance: 
significant direct effects and covariances (P-value < 0.05) are either 
red (estimate > 0) or blue (estimate < 0), while non-significant ones are 
gray-shaded. \textbf{Panel C} highlights in yellow all directed paths between 
genes SOD1 and NEFH, showing how \lstinline{SEMpath()} may help us to clarify 
and evaluate causal effects between perturbed source-target pairs, within 
an entangled cluster.}
\end{figure}

One of the goals of \pkg{SEMgraph} is to provide a set of causal interence 
tools also for users with minimal statistical expertise. To this end, we 
propose four preset strategies, implemented in the \lstinline{modelSearch()} 
function, combining \lstinline{SEMdag()}, \lstinline{SEMbap()}, and \lstinline{extendGraph()} 
functions. All strategies estimate a DAG through the adjusted (de-correlate) 
data matrix $Z$ by iteratively update DAG and $Z$ according to the 
following steps:
\begin{enumerate}
\item Initialization of $G^{(0)}$ , $Z^{(0)}$ and $\Psi^{(0)}$ with some 
suitable estimates; i.e., $G^{(0)} = \mathrm{DAG}^{(0)}$, $Z^{(0)} = Y$, 
and $\Psi^{(0)} = I$.
\item Given ($G^{(t)}$ and $Z^{(t)}$) update $Z^{(t + 1)} = Z^{(t)} 
\Psi^{-\frac{1}{2} \, (t + 1)}$ by fitting the constrained matrix 
$\Psi^{-1 \, (t + 1)}$ after d-separation testing of either 
$\mathrm{cor}(Z_j; Z_k \, | \, \mathrm{pa}(j) \, \cup \,  
\mathrm{pa}(k)) = 0$ or the $\mathrm{cor}(Z_j; Z_k \, | \, 
\mathrm{min}(Z)) = 0$ at a given \lstinline{alpha} significance level, using 
the \lstinline{SEMbap(}$G^{(t)}, Z^{(t)}$\lstinline{)} function;
\item Given ($G^{(t)}$ and $Z^{(t + 1)}$), update $G^{(t + 1)}$ estimating 
the $\mathrm{DAG}^{(t + 1)}$ via topological order of $G^{(t)}$ and edges 
penalty weighted LASSO screening at a given \emph{beta} threshold, using 
the \lstinline{SEMdag(}$G^{(t)},Z^{(t + 1)}$\lstinline{)} function;
\item Repeat steps 2 and 3 above until convergence (i.e., 
$G^{(t)} = G^{(t + 1)}$) or the Shipley\textquotesingle s global fitting 
test P-value > 0.05.
\end{enumerate}
This procedure is implemented in the \lstinline{modelSearch()} function, 
following the same syntax of \lstinline{SEMbap()} and \lstinline{SEMdag()}. 
With \lstinline{pstop = TRUE}, the algorithm can be halted when the 
Shipley\textquotesingle s test P-value > 0.05.\\
DAG estimation can be controlled through the argument \lstinline{alpha} 
(i.e., the significance level for the FDR correction), where 0 corresponds 
to no data de-correlation, and \lstinline{beta} (i.e., the LASSO coefficient 
threshold), where 0 maintains all the edges of the input graph. We suggest 
to start with \lstinline{alpha = 0.05} and \lstinline{beta = 0.1} to have a good 
balance between model adjustment and density. Then \lstinline{beta} could be 
gradually decreased (0.1 to 0) to obtain more complex models, unless the 
Shipley\textquotesingle s global fitting test P-value > 0.05. Similarly, 
argument \lstinline{alpha} can be increased up to \lstinline{0.2}. A higher 
\lstinline{alpha} level includes more hidden covariances, thus considering 
more sources of confounding, resulting in a higher data de-correlation.\\
Considering the ALS example, the model search of a DAG using the 
\lstinline{search = "basic"} procedure has the following code:
\begin{lstlisting}
R> # Model search
R> model <- modelSearch(graph = alsData$graph, data = data.npn, gnet = NULL, 
+                       d = 0, search = "basic", beta = 0.1, 
+                       alpha = 0.05, pstop = TRUE,
+                       verbose = FALSE)
\end{lstlisting}
\begin{lstlisting}[stringstyle=\color{DarkGreen},
                   keywordstyle=\color{DarkGreen},
                   commentstyle=\color{DarkGreen}]
@##  Searching for missing covariances ... 220 
 Basis set 267 of 267

     C_test          df      pvalue 
543.7724429 534.0000000   0.3753951 
Done.

RICF solver ended normally after 2 iterations 
deviance/df: 1.755445  srmr: 0.0839307 @
\end{lstlisting}
The resulting graph is shown in Figure~\ref{fig:improvedModel}A. We may 
then evaluate model perturbation using the \lstinline{SEMrun()} function, 
as shown in Figure~\ref{fig:improvedModel}B. In addition, with 
\lstinline{SEMace()} and \lstinline{SEMpath()} we can evaluate ACE, path perturbation, 
and fitting of specific directed paths between a souce-sink pair. As an 
example, Figure~\ref{fig:improvedModel}C shows in yellow all directed 
paths between genes SOD1 (Entrez ID: 6647) and NEFM (Entrez ID: 4741).
\begin{lstlisting}
R> pert <- SEMrun(model$graph, model$data, alsData$group)
R> ace <- SEMace(model$graph, model$data, alsData$group, method = "BH")
R> path <- SEMpath(model$graph, model$data, alsData$group,
+                  from = "6647", to = "4741",
+                  path = "directed",
+                  verbose = TRUE)
\end{lstlisting}

All the steps done by \lstinline{modelSearch()} are shown to standard output, and
the resulting graphs are visualized in Figure~\ref{fig:improvedModel} A-B-C.
Following the example above, the extracted DAG model  has a good fitting
(deviance/df < 2, srmr near 0.08, and C-test with P-value > 0.05). The output
\lstinline{model} object contains model fitting as a \pkg{lavaan} object
 (\lstinline{model$fit}), the output graph coloured according to node and edge
relevance during the estimation steps (\lstinline{model$graph}), and the adjusted 
dataset (\lstinline{model$data}). 
With \lstinline{search = "basic"}, we enabled a \emph{data-driven} model 
search strategy, where model structure is based only on data 
and no validation against a reference network is done (i.e., 
\lstinline{gnet = NULL} and \lstinline{d = 0}). In the example above, we set 
\lstinline{beta} to \lstinline{0.1} to reduce graph density. As a result, 
input edges could be removed and new ones could be added, partially reshaping 
model architecture. The aim is to generate an improved model, achieving 
a good overall fitting (for DAGs, the main fitting index is the 
Shipley\textquotesingle s global test P-value > 0.05), showing the best 
possible balance among model complexity, fitting, and perturbation.\\
In this example, the output model shows how the SOD1 gene deregulation 
is causally connected to the deregulated gene NEFM, implied in the 
maintainance of a physiological neuronal caliber. This indirect connection 
(the yellow path in Figure~\ref{fig:improvedModel}C), absent in the input 
model (Figure~\ref{fig:E1_ALS_groupEffect}), is now possible thanks 
to the new connections BCL2-DAXX, CYCS-MAPK13, and TOMML40-MAP2K6 (red 
links in Figure~\ref{fig:improvedModel}A), showing a tight association 
between apoptosis and neuronal caliber regulation, both dysregulated in 
neurodegenerative disorders.\\
Conversely, we could take advantage of known interactions, importing them 
in our model to extend it. We define them as \emph{knowledge-based} strategies. 
The \emph{outer} (\lstinline{search = "outer"}) strategy relies on an external 
reference network and the input graph topology, to assess the presence 
of possible hidden mediators (\lstinline{d} > 1), including them in the 
output model. If one is not interested in adding new mediators from the 
reference, but still wants to evaluate the presence of internal hidden 
indirect (i.e., mediated) paths, the \lstinline{search} argument can be set 
to \lstinline{"inner"}: the reference network is still used, but only to 
validate the new direct and indirect paths added to the model. 
Both \emph{inner} and \emph{outer} search strategies rely on the initial 
estimation of a DAG, working as a causal model backbone. Finally, 
we can use a \emph{direct} strategy (\lstinline{search = "direct"}), where 
the input graph structure is improved only through direct (i.e., adjacent) 
link search, followed by interaction validation and import from the 
reference network, with no mediators (i.e., \lstinline{d = 1}).

\section{Network clustering and scoring} \label{sec:semclust}

\begin{figure}[t!]
\centering
\includegraphics[scale=0.75]{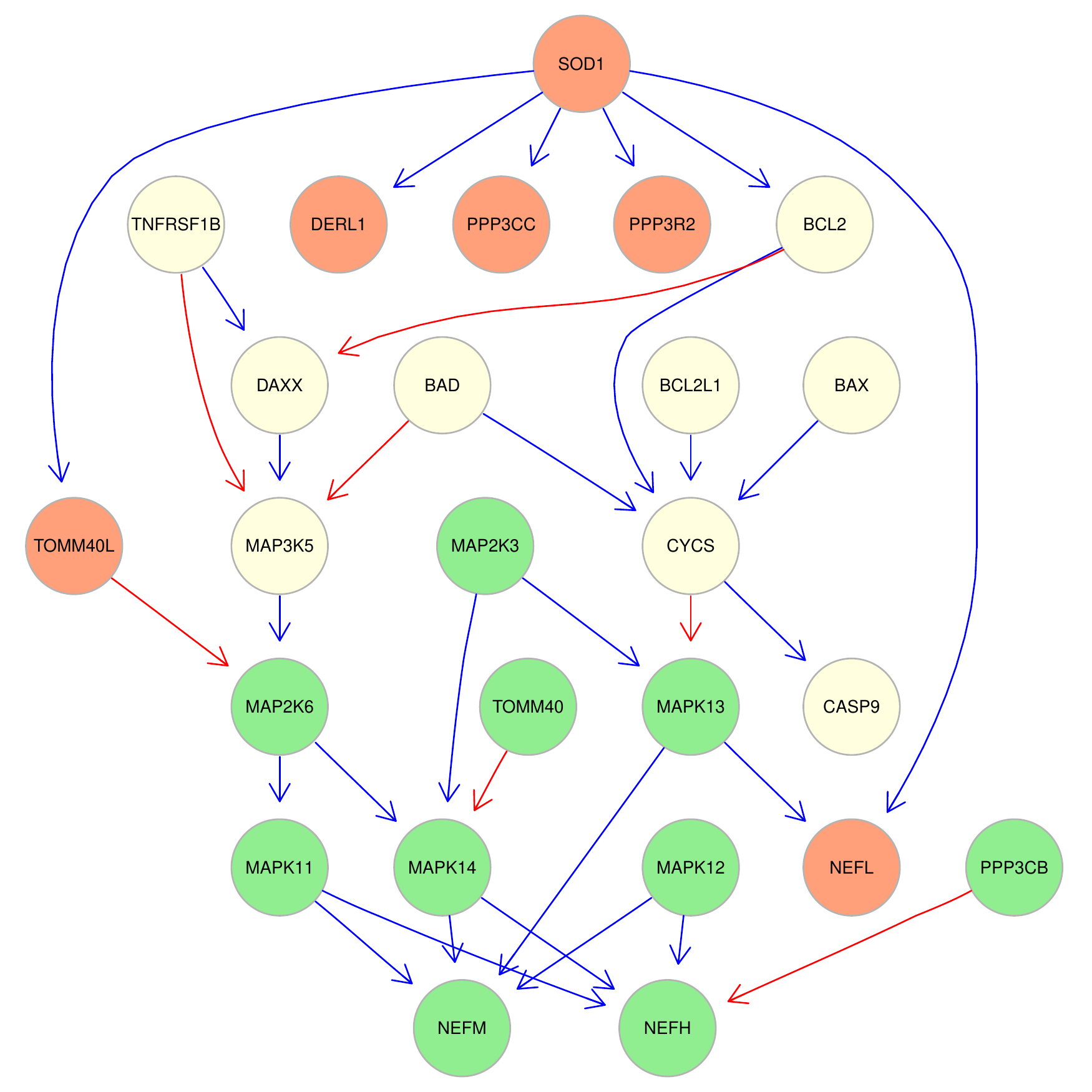}
\caption{\label{fig:ALS clustering}  Colored clustering of the ALS model using
\lstinline{clusterScore} with the edge betweeness  algorithm (\lstinline{type = "ebc"}).
Each cluster has its specific functional characterization:  SOD1 and phosphatases
module (lightsalmon), BCL2 and caspases module (lightyellow),  MAPK module (lightgreen).}
\end{figure}

\pkg{SEMgraph} offers the possibility to define topological communities 
of an input graph, generating scores for each statistical unit (i.e., 
subject) by using data from nodes belonging to communities. Clusters can 
be defined using the algorithms implemented in the \proglang{R} package 
\pkg{igraph} \citep{igraph2006} and then they can be fitted as independent 
models. Among the available clustering methods, we suggest either the walktrap 
community detection algorithm (WTC), based on random walks and develobed 
by \cite{wtc2005}, or the edge betweenness clustering (EBC), developed 
by \cite{ebc2004}. The former tends to generate as many clusters as needed 
to cover the whole input network. The latter generally produces one large 
subnetwork and other much smaller communities or singletons. In case of 
trees, our implementation of the tree agglomerative hierarchical clustering 
(TAHC), proposed by \cite{MST2015}, is the suggested solution. 
Our aim here is to provide a tool yielding different (orthogonal) local 
models when dealing with large networks ($|V| > 100$). Beside network 
size, we generally recommend clustering when there are evidences of 
possible functional modules (i.e., subnetworks whose members are involved 
in a specific process).\\
Sample scoring can be generated by three different \emph{hidden} models:
the latent variable (LV) model, the composite variable (CV) model,
and the unobserved variable(UV) model.
The LV model consists in a confirmatory factor analysis (CFA) with one 
factor and specific error variances \citep{fsr2012}:
\begin{equation} \label{eq:lvmodel}
Y_j = \lambda_j F + E_j \quad \mathrm{with} 
\quad \mathrm{var}(F) = 1 
\, \, \, \mathrm{and} \, \, \, \mathrm{var}(E_j) = \psi_j
\end{equation}
The CV model consists in a CFA with one factor and equal (common) error 
variances, equivalent to a principal component analysis (PCA) \citep{fsr2012}:
\begin{equation} \label{eq:cvmodel}
Y_j = \lambda_j C + E_j \quad \mathrm{with} 
\quad \mathrm{var}(C) = 1 
\, \, \, \mathrm{and} \, \, \, \mathrm{var}(E_j) = \psi
\end{equation}
The UV model corresponds to a fixed factor analysis (FFA) model with one 
factor projected on the observed $X$ set, with zero residual variance, 
and equal (common) error variances fixed to 1. This is equivalent to a 
reduced-rank regression analysis (RRA) \citep{rra1982}:
\begin{align} \label{eq:uvmodel}
Y_j &= \lambda_j U + E_j \quad \mathrm{with} 
\quad \mathrm{var}(U) = 1 
\, \, \, \mathrm{and} \, \, \, \mathrm{var}(E_j) = 1\\
U &= \sum_{}^{} \gamma_k X_k
\end{align}
In every hidden model, $Y_j$ are the random observed endogenous variables 
of each module, and $E_j$ the residual errors, with 
$j = (1 \mathrm{,} \, \dots \mathrm{,} \, q)$. 
In the UV model, $X_k$ represent the observed variables, with 
$k = (1 \mathrm{,} \, \dots \mathrm{,} \, r)$. 
Variables $F$, $C$, and $U$ correspond to the scores assigned to each 
subject, for each cluster, representing the latent factor, the principal 
component, and the unmeasured variable of the hidden model, respectively. 
In the UV model, the factor scores $U$ are found in the space spanned by 
the source variables $X$ of each module (i.e., they are projected on $X$). 
Factor scores $U$ are also called \emph{unmeasured variables}, rather 
than latent variables or factors, because they can be expressed as a function 
of the observed $X$ variables. Although the underlying variables are not 
actually measured, the scores $U$ are measurable \citep{SEM1980}.\\
\pkg{SEMgraph} generates cluster scores using the \lstinline{factor.analysis()} 
function of the \proglang{R} package \pkg{cate} \citep{cate2019}, an 
efficient package for high-dimensional factor analysis models. Only modules 
for which cluster scores represent 50\% or more of the total variance are 
considered. The general syntax for network clustering is the following:
\begin{lstlisting}
R> U <- clusterScore(model$graph, model$data, alsData$group, HM = "LV",
+                    type = "ebc", size = 5)
\end{lstlisting}
Arguments \lstinline{type} and \lstinline{size} set the clustering algorithm and 
the minimum group of nodes to generate a cluster (groups smaller than 
\lstinline{size} are considered as singletons). The suggested \lstinline{type} is 
the one between the walktrap (\lstinline{"wtc"}) and edge betweeness 
(\lstinline{"ebc"}) community detection algorithm resulting in the largest 
number of nodes included in clusters, with a minimum cluster size of 5. 
Argument \lstinline{HM} determines the type of \emph{hidden model} used to 
generate cluster scores: latent variable model (\lstinline{HM = "LV"}), 
composite variable model (\lstinline{HM = "CV"}), and unobserved variable 
model (\lstinline{HM = "UV"}). The global effect of the \lstinline{group} on every 
cluster can be viewed using \lstinline{parameterEstimates()}:
\begin{lstlisting}
R> scores <- parameterEstimates(U$fit)
R> head(scores)
\end{lstlisting}
\begin{lstlisting}[stringstyle=\color{DarkGreen},
                   keywordstyle=\color{DarkGreen},
                   commentstyle=\color{DarkGreen}]
@  lhs op   rhs    est    se      z pvalue ci.lower ci.upper
1 LV1  ~ group -0.471 0.238 -1.976  0.048   -0.938   -0.004
2 LV2  ~ group  0.042 0.249  0.167  0.867   -0.447    0.531
3 LV3  ~ group  0.744 0.287  2.588  0.010    0.180    1.307
4 LV1 ~~   LV1  1.037 0.116  8.944  0.000    0.809    1.264
5 LV2 ~~   LV2  1.135 0.127  8.944  0.000    0.886    1.384
6 LV3 ~~   LV3  1.507 0.168  8.944  0.000    1.176    1.837@
\end{lstlisting}
Every cluster is represented by a LV and each estimate measures the 
global effect of the group over it. Together with the fitted hidden model 
\lstinline{U$fit}, \lstinline{clusterScore()} returns the \lstinline{data.frame} containing 
cluster scores (\lstinline{U$dataHM}) and a vector indicating the cluster 
membership for every node (\lstinline{U$membership}). 
Topological cluster networks (without subject scoring) can be produced 
independently from \lstinline{clusterScore()}, using the \lstinline{clusterGraph()} 
utility:
\begin{lstlisting}
R> C <- clusterGraph(model$graph, type = "ebc", size = 5, verbose = FALSE)
\end{lstlisting}
The \lstinline{clusterGraph()} arguments are equivalent to those used in 
\lstinline{clusterScore()}. In addition, function \lstinline{cplot()} generates 
and plots separate graphs for each cluster, and returns the input graph 
with a new attribute \lstinline{V(graph)$color}, where each cluster membership 
correspond to a different color:
\begin{lstlisting}
R> G <- cplot(graph = model$graph, membership = U$membership, map = TRUE)
\end{lstlisting}
Arguments \lstinline{graph} and \lstinline{membership} correspond to the input graph 
and node membership, respectively. If the \lstinline{map} argument is set to 
\lstinline{TRUE}, the input graph is colored according to cluster membership 
(object \lstinline{G$graph}), as shown in Figure~\ref{fig:ALS clustering}.

If we consider clusters as local models, we can extract and fit them 
through the function \lstinline{extractClusters()}:
\begin{lstlisting}
R> G <- extractClusters(model$graph, model$data, alsData$group,
+                       membership = U$membership)
\end{lstlisting}
\begin{lstlisting}[stringstyle=\color{DarkGreen},
                   keywordstyle=\color{DarkGreen},
                   commentstyle=\color{DarkGreen}]
@  cluster N.nodes N.edges dev_df  srmr   pv.act   pv.inh
1     HM1       6       5  1.894 0.056 0.922706 0.043883
2     HM2      10      13  2.282 0.068 0.000000 0.011690
3     HM3       9      10  1.919 0.077 0.118250 0.108623@
\end{lstlisting}
The object \lstinline{G} contains the list of clusters as separated igraph 
objects (\lstinline{G$clusters}) and a list of fitting results (\lstinline{G$fit}). 
The summary statistics shown above are stored in the object \lstinline{G$dfs}.

\section{Network weighting and filtering} \label{sec:semWeightRed}

A common problem in network biology and medicine is to filter large 
models (i.e., networks) to highlight informative interactions, paths 
and communities, and find phenotype-associated factors. Although 
network topology alone provides enough information for many filtering 
algorithms to work, edge and node weights may significantly improve this 
process. Unlikely several network analysis tools, \pkg{SEMgraph} can work 
on both edge- and node-weighted graphs.

\subsection{Network weighting} \label{sec:graphWeight}

\pkg{SEMgraph} uses two different strategies to weight nodes and edges 
on the base of the perturbation induced by an external influence (that 
is internally represented by the binary group variable). Node-level 
perturbation consists in the detection of a subset of nodes, called 
\emph{seeds}, having a key topological or functional role. Currently three 
seed detection methods are provided: closeness percentile (\lstinline{q}), 
prototype clustering (\lstinline{h}), and t-test (\lstinline{alpha}). Three binary 
\lstinline{seed} attributes (\lstinline{1}: seed, \lstinline{0}: non-seed) are associated 
to each node. The first set includes nodes with closeness larger than 
the q-th percentile (computed through the \pkg{igraph} function 
\lstinline{closeness()}). The second set includes nodes belonging to the prototype 
cluster generated using the \proglang{R} package \pkg{protoclust} 
\citep{protoclust2011} by cutting at distance $h = 1 - |r|$ (with $r$ 
being the Pearson\textquotesingle s correlation coefficient). The third 
set of seeds includes nodes with significant group effect at \lstinline{alpha} 
level as measured by P-values testing a bivariate linear model, fitted 
with the \proglang{R} function \lstinline{lm()}.\\
Edge-level perturbation consists into edge weights (P-values 
and $z$ signs) using three different trivariate procedures: a SEM model, 
a covariance model, and the Fisher\textquotesingle s $r$-to-$z$ transform.\\
The SEM model implies testing the group effects on the source node $j$ 
and the sink node $k$. A common group effect model of $X = \{
0\mathrm{: control}; 1\mathrm{: case}\}$ is fitted:
\begin{equation} \label{eq:covmodel}
Y_j = \beta_{jk} Y_k + \beta_j X + U_j \mathrm{,} \quad 
Y_k = \alpha_k X + U_k \mathrm{,} \quad 
\mathrm{cov}(U_j; U_k) = 0
\end{equation}
and a weighted sum defines the new parameter $w$ combining the total 
effect (TE) of the binary group on source and sink nodes:
\begin{equation} \label{eq:wjk}
w_{jk} = \mathrm{abs}(\beta_j + \alpha_k \beta_{jk})/d_j + 
\mathrm{abs}(a_k)/d_k
\end{equation}
where $d_k$ and $d_j$ are the outgoing degrees (i.e., the number of all 
direct outgoing connections in the input graph, for each node) of the 
$k$-th sources and $j$-th sinks, respectively. The P-values are computed 
through the z-test = $w/\mathrm{SE}(w)$ of the combined TE of the group 
on the source node $j$ to the sink node $k$.\\
The covariance model implies testing the group effect simultaneously on 
the source node $j$, the sink node $k$ and their interaction 
$(j; k)$. In this case, a two-group (0: controls, 1: cases) 
covariance model with an intercept parameters is fitted:
\begin{align} \label{eq:covm2gr}
Y_{j}^{(1)} &= \beta_j + U_{j}^{(1)} \mathrm{,} \quad 
Y_{k}^{(1)} = \beta_k + U_{k}^{(1)} \mathrm{,} \quad 
\mathrm{cov}(Y_{j}^{(1)}; Y_{k}^{(1)}) = \psi_{jk} \quad \\
Y_{j}^{(0)} &= \alpha_j + U_{j}^{(0)} \mathrm{,} \quad 
Y_{k}^{(0)} = \alpha_k + U_{k}^{(0)} \mathrm{,} \quad 
\mathrm{cov}(Y_{j}^{(0)}; Y_{k}^{(0)}) = \phi_{jk} \quad
\end{align}
A weighted sum defines the new weight parameter $w$, combining the group 
effect on the source node (mean difference $\beta_j - \alpha_j$), the sink 
node (mean difference $\beta_k - \alpha_k$), and the source-sink connection 
(covariance difference $\psi_{jk} - \phi_{jk}$):
\begin{equation} \label{eq:wjk2gr}
w_{jk} = \mathrm{abs}(\beta_j - \alpha_j)/d_j + 
\mathrm{abs}(\beta_k - \alpha_k)/d_k + 
\mathrm{abs}(\psi_{jk} - \phi_{jk})
\end{equation}
where $d_k$ and $d_j$ are the degree (i.e., the number of incoming and 
outgoing connections) of the source node $k$ and sink node $j$, 
respectively. P-values are computed throug the t-test = $w/\mathrm{SE}(w)$ 
on the combined difference of the group over the source node $j$, the 
sink node $k$, and their connection $(j; k)$.\\
Finally, the correlation method tests the group difference between the 
correlation coefficients $r_{jk}^{(1)}$ and $r_{jk}^{(0)}$ of connected 
nodes $j$ and $k$, by applying the Fisher\textquotesingle s $r$-to-$z$ 
transform: $z = 0.5 \, \mathrm{log}[(1 + r)/(1 - r)]$, with 
$t = (z^{(1)} - z^{(0)}) / \sqrt{1/(n_1 - 3) + 1/(n_0 - 3)}$ \citep{r2z1915}. 
Usually, the SEM model is similar to the covariance model if the 
covariance difference is close to zero. The correlation method is similar 
to the covariance model if the source and sink mean differences are close 
to 0. From a computational point of view, the $r$-to-$z$ transform is 
the fastest of the three methods. The $\mathrm{SE}(w)$ is estimated through 
\pkg{lavaan}, specifying the new parameter $w$ using the \lstinline{:=} 
operator.\\
Graph weighting can be applied to any graph with the \lstinline{weightGraph()} 
function:
\begin{lstlisting}
R> W <- weightGraph(alsData$graph, alsData$exprs, alsData$group,
+                   method = "r2z", seed = c(0.05, 0.5, 0.5))
\end{lstlisting}
Here we used the \lstinline{group} and \lstinline{seed} arguments to include the 
perturbation information (i.e., the case-control difference) into weights 
(by default, \lstinline{seed = "none"}). 
The object \lstinline{W} correspond to the input graph with three new node 
(i.e., seed) attributes (namely, \lstinline{V(W)$pvlm}, \lstinline{V(W)$proto}, and 
\lstinline{V(W)$qi}) and two new edge attributes (namely, \lstinline{E(W)$pv} and 
\lstinline{E(W)$zsign}). Each of the three seed attributes is a binary vector, 
taking value \lstinline{1} for a seed and \lstinline{0} for a non-seed. Each seed 
type can be defined by a vector of three cutoffs: the significance level 
of the direct group effect, the prototype clustering 
distance corresponding to $1 - \mathrm{abs}(r)$ (with $r$ being the 
Pearson\textquotesingle s correlation coefficient), and the closeness 
percentile, respectively equal to 0.05, 0.5 and 0.5 in the above \lstinline{seed} code.
The edge \lstinline{pv} attribute is the vector of P-values yielded by 
the selected method (in the example above, we used \lstinline{"r2z"}), providing 
continuous edge weights. Methods \lstinline{"sem"} and \lstinline{"cov"} can be 
specified for using the SEM or covariance weighting model, respectively. 
These two methods are slower than \lstinline{"r2z"}, but multicore usage is 
automatically enabled for large networks.The edge attribute \lstinline{zsign} 
is the sign of the test statistic $z$, that can be interpreted as activated 
(+1), inhibited (-1), or neutral (0, with P-value > 0.05), providing 
categorical edge weights.

\subsection{Active module finding} \label{sec:semred}

Reducing large complex graphs by either extracting critical relationships 
or perturbed disease modules is key to focus relevant information into 
simpler subgraphs. Usually the detection of critical nodes and edges and 
disease modules supplements prior knowledge about disease-associated 
genomic elements, leveraging on emergent properties that can be revealed 
only through network analysis. Although a wide range of different methods 
have been proposed through years \citep{Network2020}, \pkg{SEMgraph} 
proposes four fast procedures, including: random walk with restart (RWR), 
heat diffusion (HDI) \citep{diffusr2018}, Steiner tree (ST) 
\citep{kou1981}, and the union of shortest path graph (USPG) 
\citep{Shortest2010}.\\
The RWR and HDI algorithms, implemented in the \proglang{R} package 
\pkg{diffusR} \citep{diffusr2018}, starts from an initial distribution 
of node P-values, then computing their stationary distribution. They spread 
information in the form of node weights along the edges of a graph to other 
nodes. The information (i.e., node weights) is iteratively propagated to 
other nodes until an equilibrium state or stop criterion occurs. The RWR 
starts from an initial distribution $p_0$, then computing their stationary 
distribution. The process depends on a restart probability parameter 
that allows to regulate how often the RWR returns back to the initial 
values. The HDI starts from an amount of heat $h_0$, and gets stationary 
distribution using the Laplacian matrix of a graph. Every iteration (or 
time interval) $t$ heat streams from the starting nodes into surrounding 
nodes. The reduce graph $G_0$ for both RWR and HDI is the induced subgraph 
of the input graph $G$ defined by the $S$ nodes in the top-rank scoring 
($q$-th quantile of the stationary distribution).\\
The ST and USPG algorithms minimize paths between seed nodes $S$ with 
minimum weights (i.e., maximum perturbation), we will refer to these paths 
as the maximum perturbation paths (MPPs), and a perturbation route being 
an MPP subset. Edge weights are defined as inverse of negative logarithm 
of the P-values. In this way, edge weights are in a positive continuous 
range $[0, \infty)$. We refer to this scale as perturbance 
\citep{ItalianFTD2017}. The lower the P-value (or the $w$-value), the 
higher the perturbation. The ST problem \citep{kou1981} is to find a 
connected subgraph $G_0$ of $G$ such that the additional nodes $C$ 
(called the \emph{Steiner} or \emph{connector} nodes) connecting \emph{a 
priori} (e.g., disease) seed nodes $S$ (called the \emph{terminal} nodes) 
minimize the sum of the weight of every edge in the subgraph $G_0$. 
Various heuristic algorithms are available for solving ST problem. 
The \pkg{SEMgraph} function \lstinline{SteinerTree()} applies a fast 
algorithm approximation, first proposed by \cite{kou1981}.
\lstinline{USPG()} generates a subgraph $G_0$ of $G$ as the union of the 
significantly pertubed shortest paths between \emph{a priori} seed nodes 
$S$. The USPG problem is focused in computing the minimum shortest path 
between each pair of nodes, selecting MPPs. Thus, each shortest path 
considers not only the number of links needed to reach the disease-associated 
node, but also the number of disease-associated edges that are included 
in the path.\\
Active module finding methods are implemented in the wrapper 
\lstinline{activeModule()}:
\begin{lstlisting}
R> R <- activeModule(W, type = "rwr", seed = "pvlm", eweight = "pvalue")
\end{lstlisting}
Function \lstinline{activeModule()} takes two main inputs: a weighted network 
\lstinline{W} and a reduction method specified by the argument \lstinline{type}, 
including algorithms: RWR (\lstinline{"rwr"}), HDI (\lstinline{"hdi"}), the 
Kou version of the Steiner tree algorithm (\lstinline{"kou"}), and the USPG 
method (\lstinline{"usp"}). The optional argument \lstinline{seed} takes either a 
binary vector of custom seeds or one among \lstinline{"qi"}, \lstinline{"proto"}, 
or \lstinline{"pvlm"}, corresponding to the seed types generated through the 
\lstinline{weightGraph()} function. Finally, the optional argument \lstinline{eweight} 
allow the user to specify either a custom vector of distances or one between 
\lstinline{"pvalue"} and \lstinline{"zsign"}, generated using the \lstinline{weightGraph()} 
function.

\section{Graph conversion utilities} \label{sec:semconvert}

\pkg{SEMgraph} uses two standard graph and model formats, being respectively 
\lstinline{igraph} and \lstinline{lavaan}. Different \pkg{SEMgraph} functions may 
require DAG conversion, and a common way of generating undirected graphs 
from data is to convert a correlation matrix to an adjacency matrix, using 
a threshold over the correlation coefficient. These conversion types can 
be obtained with the following functions:
\begin{lstlisting}
R> # SEM (lavaan syntax) from an igraph object
R> G <- model$graph
R> V(G)$name <- mapIds(org.Hs.eg.db, V(G)$name, 'SYMBOL', 'ENTREZID')
R> als.sem <- graph2lavaan(G)

R> # igraph network object from SEM (lavaan syntax)
R> als.graph <- lavaan2graph(als.sem, directed = TRUE, psi = FALSE)

R> # Extract a DAG from a network (igraph format)
R>  i <- which(names(kegg.pathways)=="Notch signaling pathway")
R> G<- properties(kegg.pathways[[i]])[[1]]
R> dag <- graph2dag(G, data = data.npn, bap = FALSE)
R> bap <- graph2dag(G, data = data.npn, bap = TRUE)

R> # Extract an undirected network from a correlation matrix R
R> R <- cor(model$data)
R> U <- corr2graph(R, n = nrow(model$data), type = "marg", method = "BH", 
+                  alpha = 0.05)
\end{lstlisting}
Function \lstinline{graph2lavaan()} simply generates a SEM (lavaan syntax) 
from the input igraph object, while \lstinline{lavaan2graph()} does the 
opposite operation. Natively, path diagrams are directed and may contain 
covariances (i.e., bidirected edges). However, lavaan2graph() may generate 
either an undirected graph (\lstinline{directed = FALSE}) or a directed graph 
without covariances (\lstinline{psi = FALSE}). By default, both these arguments 
are set to \lstinline{TRUE}.\\
Function \lstinline{graph2dag()} extract a DAG from an input graph through a
two-steps pruning strategy. Firstly, bidirected edges are removed from the
input graph. Secondly, edge are weighted by P-values, through marginal
correlation testing. When a cycle is detected,  the edge with highest P-value
is removed, breaking the cycle. If \lstinline{bap=TRUE}, a BAP is then generated 
merging the output DAG and the bidirected edges from the input graph.\\
The function \lstinline{corr2graph()} offers the possibility to apply a threshold 
to the input correlation matrix (\lstinline{R}), based on either marginal 
(\lstinline{type = "marg"}) or conditional (\lstinline{type = "cond"}) correlation 
testing. The arguments alpha and method set the significance level over 
the adjusted P-value. In addition, the Minimum Spanning Tree 
(\lstinline{type = "mst"}) or the Triangulated Maximally Filtered Graph 
(\lstinline{type = "tmfg"}) are implemented for filtering the amount of meaningful 
correlation structure. A MST is a subset $G = (V = p, E = p - 1)$ of a 
edge-weighted graph that connecting all the $p$ nodes (variables) together, 
without cycles and with the minimum possible total edge weight. The TMFG 
method \citep{TMFG2016} uses a structural constraint that limits the number 
of zero-order correlations included in the network, yielding the subgraph 
$G = (V = p, E = 3p - 6)$.

\section{Disease modules detection} \label{sec:semlarge}

In this case study, we want to build a causal model for the Frontotemporal 
Dementia, a neurodegenerative disorder characterized by cognitive and 
behavioural impairments \citep{ItalianFTD2017}. The aim is to produce a 
map of the DNA methylation (DNAme) alterations caused by FTD, without an 
initial disease model. For this example, we will use DNAme data from 
\cite{FTD2015} (GEO accession: GSE53740), stored in the \pkg{SEMdata} 
package. Although not necessary, having a collection of known 
disease-associated networks is an advantageous starting point. For instance, 
the KEGG BRITE database allows to search for terms, including human disorders, 
that could be associated to one or more pathways. The term Frontotemporal 
lobar degeneration (an alias for FTD; KEGG ID: H00078) is associated to 
6 KEGG pathways: \emph{MAPK signaling pathway} (hsa04010), \emph{Protein 
processing in endoplasmic reticulum} (hsa04141), \emph{Endocytosis} 
(hsa4144), \emph{Wnt signaling pathway} (hsa04310), \emph{Notch signaling 
pathway} (hsa04330), and \emph{Neurotrophin signaling pathway} (hsa04722). 
Starting from database queries is not a requirement. Initial network models 
may also derive from exploratory analyses, such as overrepresentation 
analysis (ORA) or gene set enrichment analysis (GSEA) \citep{enrichment2019}.
In \pkg{SEMgraph}, we can use the \lstinline{SEMgsa()} utility to apply gene 
set analysis (GSA) on a collection of networks.
\begin{lstlisting}
R> # load libraries
R> library(SEMdata)
R> library(huge)

R> # FTD-related pathway selection
R> ftd.pathways <- c("MAPK signaling pathway",
+                    "Protein processing in endoplasmic reticulum",
+                    "Endocytosis",
+                    "Wnt signaling pathway",
+                    "Notch signaling pathway",
+                    "Neurotrophin signaling pathway")
R> j <- which(names(kegg.pathways) %in% ftd.pathways)

R> # Nonparanormal transform of DNAme PC1 data
R> pc1.npn <- huge.npn(ftdDNAme$pc1)

R> # Gene set analysis (GSA)
R> ftd.gsa <- SEMgsa(kegg.pathways[j], pc1.npn, ftdDNAme$group, n_rep = 5000)

R> # Input graph as the union of FTD KEGG pathways
R> graph <- graph.union(kegg.pathways[j])
R> graph <- properties(graph)[[1]]

R> # Seed extraction
R> seed <- V(graph)$name[V(graph)$name %in% unique(unlist(ftd.gsa$DRN))]
\end{lstlisting}
This leads to a graph of 581 nodes and 3817 edges, and a seed list of 
29 DRNs. With large networks, a heuristic solution to maximize model 
perturbation within a relatively less complex model is graph weighting 
and filtering. Fisher\textquotesingle s $r$-to-$z$ is the fastest 
weighting solution for very large networks:
\begin{lstlisting}
R> W <- weightGraph(graph, pc1.npn, ftdDNAme$group, method = "r2z")
\end{lstlisting}
We may then apply \lstinline{activeModule()} to generate our reduced perturbed 
model. A very fast and efficient solution to estimate the perturbed 
backbone of the input graph is the Steiner tree (\lstinline{type = "kou"}), 
traversing all our seeds, while minimizing the total weight of the network
(i.e., maximizing edge perturbation):
\begin{lstlisting}
R> # Steiner tree extraction
R> R <- activeModule(W, type = "kou", seed = seed, eweight = "pvalue")

R> # Entrez ID conversion
R> V(R)$label <- mapIds(org.Hs.eg.db, V(R)$name, 'SYMBOL', 'ENTREZID')

R> # Perturbation evaluation and plotting
R> pert <- SEMrun(graph = R, data = pc1.npn, group = ftdDNAme$group)
R> gplot(pert$graph)
\end{lstlisting}
\begin{figure}[t!]
\centering
\includegraphics[scale=0.35]{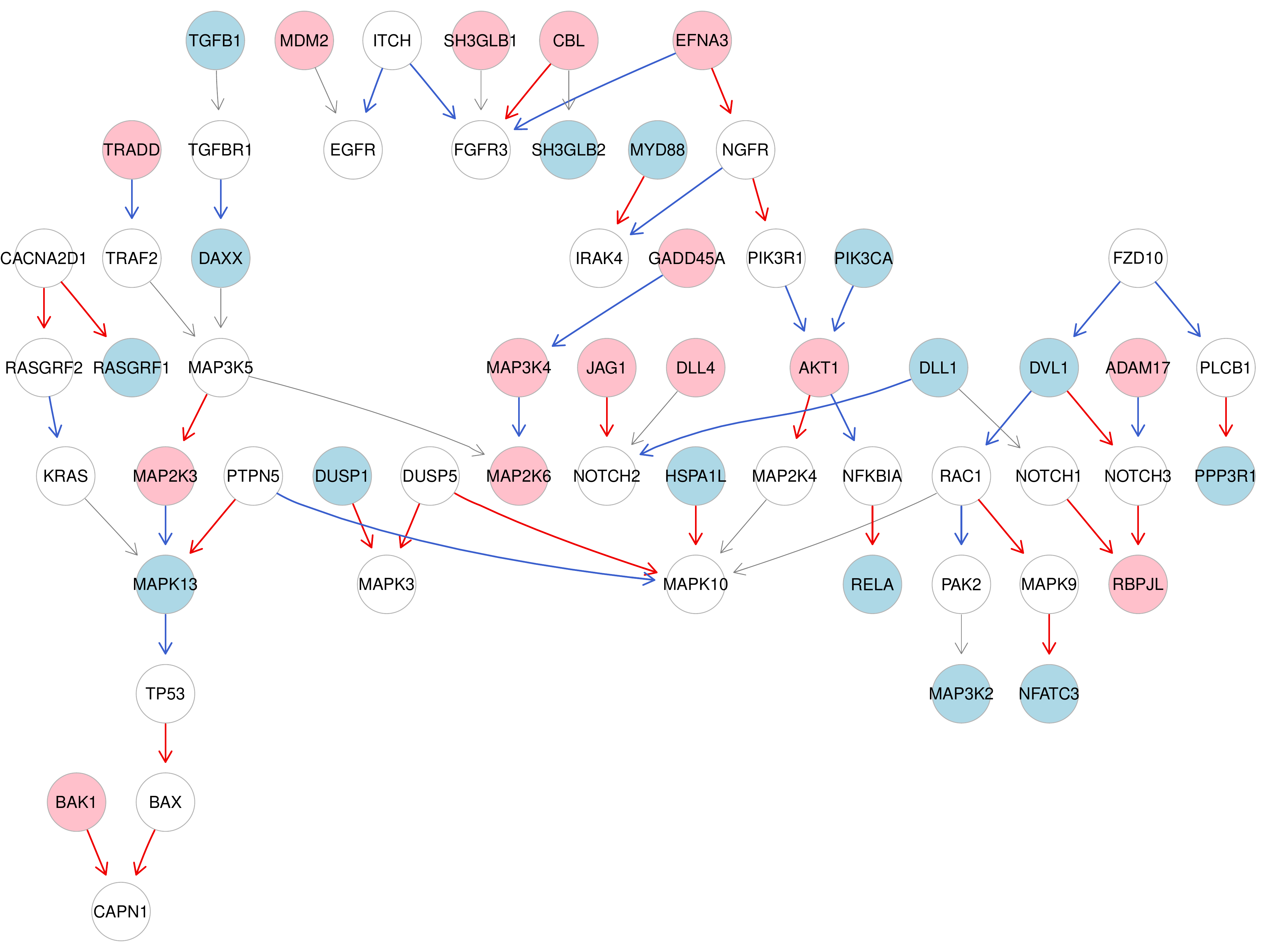}
\caption{\label{fig:FTDactiveModule} FTD perturbed backbone estimated 
from DNA methylation data. This perturbed backbone is extracted from 
the weighted input graph, maximizing both edge perturbation (i.e., 
minimizing the total weight of the tree) while traversing all the 
seeds (i.e., the Steiner connectors), defined as the nodes that are 
significantly perturbed by the diseased phenotype. The 29 seeds here 
reported were calculated using \lstinline{SEMgsa()}. Node and edge color 
coding follow the same rules applied in figure \ref{fig:improvedModel}B.}
\end{figure}
The \lstinline{kou} algorithm yielded a Steiner tree \lstinline{R} of 59 nodes and 
58 edges. In case of very large and dense network, Steiner trees are fast 
and accurate solutions for finding the essential backbone of the network, 
providing a valuable insight of the key mediators supporting the information 
flow of the system. The inferred FTD perturbed backbone, given DNAme data, 
is shown in Figure~\ref{fig:FTDactiveModule}. The perturbed backbone 
(i.e., the tree connecting seeds, maximizing edge perturbation, with 
minimum possible cost), can be exploited to build an improved causal model 
with the \lstinline{modelSearch()} function (see Section~\ref{sec:semStrategies}). 
Code and output of the backbone improvement pipeline can be found in 
the supplementary file available at: 
\url{https://github.com/fernandoPalluzzi/SEMgraph/blob/master/SEMgraph-replicationCode.R}.

\section{Summary and discussion} \label{sec:summary}

\pkg{SEMgraph} is a fast and user-friendly, yet powerful package for 
causal network analysis. Bridging graph theory and structural equation 
modeling (SEM), it conveys causal structure learning within the framework 
of multivariate linear networks, combining accurate data-driven discovery 
and confounding adjustment to model interpretability. The \pkg{SEMgraph} 
philosophy includes two main aspects: the technical aspect of usability 
and the concept of contextual analysis. The former is achieved by introducing 
automated and data-driven settings for both optimal algorithm tuning and 
scalability, to relieve the user from time-consuming and/or arbitrary choices. 
The latter is founded on the notion of model fitting/perturbation tradeoff, 
implying that destabilizing or pathological processes arise within a stable 
or phyisiological system context, under the action of a perturbing signal. 
Given the advance in causal structure learning, our direction is to 
incorporate the most recent proposals in DAG search \citep{Causality2018}, 
cyclic SEM fitting \citep{SEM2019}, and confounding adjustment 
\citep{Deconfounding2020}. In addition, new examples, pathways and 
interactome for graph usability can be easily added in \pkg{SEMdata}, 
providing an extensible platform.

\section*{Computational details}

The results in this paper were obtained using
\proglang{R}~4.1.2 with the \pkg{SEMgraph}~1.0.5 package, available at 
\url{https://github.com/fernandoPalluzzi}. All datasets are provided in 
the \pkg{SEMdata}~1.0.4 package, available at the same website.
\proglang{R} itself and all used packages are available from the Comprehensive 
\proglang{R} Archive Network (CRAN) at \url{https://CRAN.R-project.org} 
or the Bioconductor repository at \url{https://bioconductor.org}.


\bibliographystyle{unsrtnat}
\bibliography{semrefs}

\begin{thebibliography}{76}
\newcommand{\enquote}[1]{``#1''}
\providecommand{\natexlab}[1]{#1}
\providecommand{\url}[1]{\texttt{#1}}
\providecommand{\urlprefix}{URL }
\expandafter\ifx\csname urlstyle\endcsname\relax
  \providecommand{\doi}[1]{doi:\discretionary{}{}{}#1}\else
  \providecommand{\doi}{doi:\discretionary{}{}{}\begingroup
  \urlstyle{rm}\Url}\fi
\providecommand{\eprint}[2][]{\url{#2}}

\bibitem[{Akaike(1974)}]{AIC1974}
Akaike H (1974).
\newblock \enquote{A new look at the statistical model identification.}
\newblock \emph{IEEE Transactions on Automatic Control}, \textbf{19}(6),
  716--723.
\newblock \doi{10.1109/TAC.1974.1100705}.

\bibitem[{Bai and Li(2012)}]{fsr2012}
Bai J, Li K (2012).
\newblock \enquote{Statistical analysis of factor models of high dimension.}
\newblock \emph{The Annals of Statistics}, \textbf{40}(1), 436--465.
\newblock \doi{10.1214/11-AOS966}.

\bibitem[{Barab\'asi \emph{et~al.}(2011)Barab\'asi, Gulbahce, and
  Loscalzo}]{NetworkMed2011}
Barab\'asi AL, Gulbahce N, Loscalzo J (2011).
\newblock \enquote{Network Medicine: A Network-based Approach to Human
  Disease.}
\newblock \emph{Nature Review Genetics}, \textbf{12}(1), 56--68.
\newblock \doi{10.1038/nrg2918}.

\bibitem[{Bentler(2016)}]{EQS6}
Bentler PM (2016).
\newblock \emph{EQS 6 Structural Equations Program Manual}.
\newblock Encino, CA.
\newblock \urlprefix\url{http://www.mvsoft.com/}.

\bibitem[{Bentler and Weeks(1980)}]{SEM1980}
Bentler PM, Weeks DG (1980).
\newblock \enquote{Linear structural equations with latent variables.}
\newblock \emph{Psychometrika}, \textbf{45}(3), 289--308.
\newblock \doi{doi.org/10.1007/BF02293905}.

\bibitem[{Bien and Tibshirani(2011)}]{protoclust2011}
Bien J, Tibshirani R (2011).
\newblock \enquote{Hierarchical Clustering With Prototypes via Minimax
  Linkage.}
\newblock \emph{J Am Stat Assoc.}, \textbf{106}(495), 1075--1084.
\newblock \doi{10.1198/jasa.2011.tm10183}.

\bibitem[{Boker \emph{et~al.}(2011)Boker, Neale, Maes, Wilde, Spiegel, Brick,
  Spies, Estabrook, Kenny, Bates, Mehta, and Fox}]{OpenMx2011}
Boker S, Neale M, Maes H, Wilde M, Spiegel M, Brick T, Spies J, Estabrook R,
  Kenny S, Bates T, Mehta P, Fox J (2011).
\newblock \enquote{OpenMx: An Open Source Extended Structural Equation Modeling
  Framework.}
\newblock \emph{Psychometrika}, \textbf{76}, 306--317.
\newblock \doi{10.1007/s11336-010-9200-6}.

\bibitem[{Bollen(1989)}]{SEM1989}
Bollen KA (1989).
\newblock \emph{Structural Equations with Latent Variables}.
\newblock 1st edition. John Wiley \& Sons, Hoboken, NJ, USA.

\bibitem[{Bollen and Stine(1992)}]{Bootstrap1992}
Bollen KA, Stine RA (1992).
\newblock \enquote{Bootstrapping Goodness-of-Fit Measures in Structural
  Equation Models.}
\newblock \emph{Sociological Methods \& Research}, \textbf{21}(2), 205--229.
\newblock \doi{10.1177/0049124192021002004}.

\bibitem[{Brito and Pearl(2002)}]{SEM2002}
Brito C, Pearl J (2002).
\newblock \enquote{A New Identification Condition for Recursive Models With
  Correlated Errors.}
\newblock \emph{Structural Equation Modeling}, \textbf{9}(4), 459--474.
\newblock \doi{10.1207/S15328007SEM0904_1}.

\bibitem[{Brown(1975)}]{Brown1975}
Brown MB (1975).
\newblock \enquote{A Method for Combining Non-Independent, One-Sided Tests of
  Significance.}
\newblock \emph{Biometrics}, \textbf{31}(4), 987--992.
\newblock \doi{10.2307/2529826}.

\bibitem[{Buhlmann and Cevid(2020)}]{Deconfounding2020}
Buhlmann P, Cevid D (2020).
\newblock \enquote{Deconfounding and Causal Regolarisation for Stability and
  External validity.}
\newblock \emph{International Statistical Review}, \textbf{88}(S1), S114--S134.
\newblock \doi{10.1111/insr.12426}.

\bibitem[{Cai \emph{et~al.}(2013)Cai, Bazerque, and Giannakis}]{SparseSEM2013}
Cai X, Bazerque JA, Giannakis GB (2013).
\newblock \enquote{Inference of Gene Regulatory Networks With Sparse Structural
  Equation Models Exploiting Genetic Perturbations.}
\newblock \emph{PLoS Computational Biology}, \textbf{9}(5), e1003068.
\newblock \doi{10.1371/journal.pcbi.1003068}.

\bibitem[{Chan(2010)}]{Shortest2010}
Chan TM (2010).
\newblock \enquote{More Algorithms for All-Pairs Shortest Paths in Weighted
  Graphs.}
\newblock \emph{SIAM Journal on Computing}, \textbf{39}(5), 2075--2089.
\newblock \doi{10.1137/08071990X}.

\bibitem[{Chen \emph{et~al.}(2019)Chen, Drton, and Wang}]{EqVarDAG2019}
Chen W, Drton M, Wang YS (2019).
\newblock \enquote{On Causal Discovery with an Equal-Variance Assumption.}
\newblock \emph{Biometrika}, \textbf{106}(4), 973--980.
\newblock \doi{10.1093/biomet/asz049}.

\bibitem[{Cooper-Knock \emph{et~al.}(2015)Cooper-Knock, Bury, Heath, Wyles,
  Higginbottom, Gelsthorpe, Highley, Hautbergue, Rattray, Kirby, and
  Shaw}]{ALS2015}
Cooper-Knock J, Bury JJ, Heath PR, Wyles M, Higginbottom A, Gelsthorpe C,
  Highley JR, Hautbergue G, Rattray M, Kirby J, Shaw PJ (2015).
\newblock \enquote{C9ORF72 GGGGCC Expanded Repeats Produce Splicing
  Dysregulation which Correlates with Disease Severity in Amyotrophic Lateral
  Sclerosis.}
\newblock \emph{PLoS One}, \textbf{10}(5), e0127376.
\newblock \doi{10.1371/journal.pone.0127376}.

\bibitem[{Csardi and Nepusz(2006)}]{igraph2006}
Csardi G, Nepusz T (2006).
\newblock \enquote{The igraph software package for complex network research.}
\newblock \emph{InterJournal}, \textbf{Complex Systems}, 1695.
\newblock \urlprefix\url{http://igraph.sf.net}.

\bibitem[{Davies and Tso(1982)}]{rra1982}
Davies P, Tso M (1982).
\newblock \enquote{Procedures for Reduced-Rank Regression.}
\newblock \emph{Journal of the Royal Statistical Society. Series C (Applied
  Statistics)}, \textbf{31}(3), 141--145.
\newblock \doi{10.2307/2347998}.

\bibitem[{Dirmeier(2018)}]{diffusr2018}
Dirmeier S (2018).
\newblock \emph{diffusr: Network Diffusion Algorithms}.
\newblock R package version 0.1.4,
  \urlprefix\url{https://CRAN.R-project.org/package=diffusr}.

\bibitem[{Drton \emph{et~al.}(2009)Drton, Eichler, and Richardson}]{RICF2009}
Drton M, Eichler M, Richardson TS (2009).
\newblock \enquote{Computing Maximum Likelihood Estimated in Recursive Linear
  Models with Correlated Errors.}
\newblock \emph{Journal of Machine Learning Research}, \textbf{10}(81),
  2329--2348.
\newblock \doi{10.1145/1577069.1755864}.

\bibitem[{Drton \emph{et~al.}(2019)Drton, Fox, and Wang}]{SEM2019}
Drton M, Fox C, Wang YS (2019).
\newblock \enquote{Computation of maximum likelihood estimates in cyclic
  structural equation models.}
\newblock \emph{The Annals of Statistics}, \textbf{47}(2), 663--690.
\newblock \doi{10.1214/17-AOS1602}.

\bibitem[{Drton \emph{et~al.}(2011)Drton, Foygel, and Sullivant}]{Identify2011}
Drton M, Foygel D, Sullivant S (2011).
\newblock \enquote{Global Identifiability of Linear Structural Equation
  Models.}
\newblock \emph{The Annals of Statistics}, \textbf{39}(2), 865--886.
\newblock \doi{10.1214/10-AOS859}.

\bibitem[{Drton and Maathuis(2017)}]{graphLearn2017}
Drton M, Maathuis MH (2017).
\newblock \enquote{Structure Learning in Graphical Modeling.}
\newblock \emph{Annual Review of Statistics and Its Application},
  \textbf{4}(1), 365--393.
\newblock \doi{10.1146/annurev-statistics-060116-053803}.

\bibitem[{Finos \emph{et~al.}(2018)Finos, Klinglmueller, Basso, Solari,
  Benetazzo, Goeman, and Rinaldo}]{flip2018}
Finos L, Klinglmueller F, Basso D, Solari A, Benetazzo L, Goeman J, Rinaldo M
  (2018).
\newblock \emph{flip: Multivariate Permutation Tests}.
\newblock R package version 2.5.0,
  \urlprefix\url{https://CRAN.R-project.org/package=flip}.

\bibitem[{Fisher(1915)}]{r2z1915}
Fisher RA (1915).
\newblock \enquote{Frequency Distribution of the Values of the Correlation
  Coefficient in Samples from an Indefinitely Large Population.}
\newblock \emph{Biometrika}, \textbf{10}(4), 507--521.
\newblock \doi{10.2307/2331838}.

\bibitem[{Fox(2006)}]{sem2006}
Fox J (2006).
\newblock \enquote{Structural Equation Modeling With the sem Package in R.}
\newblock \emph{Structural Equation Modeling}, \textbf{13}(3), 465--486.
\newblock \doi{10.1207/s15328007}.

\bibitem[{Friedman \emph{et~al.}(2010)Friedman, Hastie, and
  Tibshirani}]{Regularization2010}
Friedman J, Hastie T, Tibshirani R (2010).
\newblock \enquote{Regularization Paths for Generalized Linear Models via
  Coordinate Descent.}
\newblock \emph{Journal of Statistical Software}, \textbf{33}(1), 1--22.
\newblock \doi{10.18637/jss.v033.i01}.

\bibitem[{Grotzinger \emph{et~al.}(2019)Grotzinger, Rhemtulla, de~Vlaming,
  Ritchie, Mallard, Hill, Ip, Marioni, McIntosh, Deary, Koellinger, Harden,
  Nivard, and Tucker-Drob}]{GenomicSEM2019}
Grotzinger AD, Rhemtulla M, de~Vlaming R, Ritchie SJ, Mallard TT, Hill WD, Ip
  HF, Marioni RE, McIntosh AM, Deary IJ, Koellinger PD, Harden KP, Nivard MG,
  Tucker-Drob EM (2019).
\newblock \enquote{Genomic Structural Equation Modelling Provides Insights Into
  the Multivariate Genetic Architecture of Complex Traits.}
\newblock \emph{Nature Human Behaviour}, \textbf{3}(5), 513--525.
\newblock \doi{10.1038/s41562-019-0566-x}.

\bibitem[{Heinze-Deml \emph{et~al.}(2018)Heinze-Deml, Maathuis, and
  Meinshausen}]{Causality2018}
Heinze-Deml C, Maathuis MH, Meinshausen N (2018).
\newblock \enquote{Causal Structure Learning.}
\newblock \emph{Annual Review of Statistics and Its Application},
  \textbf{5}(1), 371--391.
\newblock \doi{10.1146/annurev-statistics-031017-100630}.

\bibitem[{Hu and Bentler(1999)}]{GOF1999}
Hu L, Bentler PM (1999).
\newblock \enquote{Cutoff Criteria for Fit Indexes in Covariance Structure
  Analysis: Conventional Criteria Versus New Alternatives.}
\newblock \emph{Structural Equation Modeling: A Multidisciplinary Journal},
  \textbf{6}(1), 1--55.
\newblock \doi{10.1080/10705519909540118}.

\bibitem[{Huang(2018)}]{lslx2018}
Huang PH (2018).
\newblock \enquote{lslx: Semi-Confirmatory Structural Equation Modeling via
  Penalized Likelihood.}
\newblock \emph{Journal of Statistical Software}, \textbf{93}(7).
\newblock \doi{10.18637/jss.v093.i07}.

\bibitem[{Jacobucci \emph{et~al.}(2016)Jacobucci, Grimm, and
  McArdle}]{regsem2016}
Jacobucci R, Grimm KJ, McArdle JJ (2016).
\newblock \enquote{Regularized Structural Equation Modeling.}
\newblock \emph{Structural equation modeling: a multidisciplinary journal},
  \textbf{23}(4), 555--566.
\newblock \doi{10.1080/10705511.2016.1154793}.

\bibitem[{Jankov\'a and van~de Geer(2015)}]{invcov2015}
Jankov\'a J, van~de Geer S (2015).
\newblock \enquote{Confidence intervals for high-dimensional inverse covariance
  estimation.}
\newblock \emph{Electronic Journal of Statistics}, \textbf{9}(1), 1205--1229.
\newblock \doi{10.1214/15-EJS1031}.

\bibitem[{Jassal \emph{et~al.}(2020)Jassal, Matthews, Viteri, Gong, Lorente,
  Fabregat, Sidiropoulos, Cook, Gillespie, Haw, Loney, May, Milacic, Rothfels,
  Sevilla, Shamovsky, Shorser, Varusai, Weiser, Wu, Stein, Hermjakob, and
  D'Eustachio}]{Reactome2020}
Jassal B, Matthews L, Viteri G, Gong C, Lorente P, Fabregat A, Sidiropoulos K,
  Cook J, Gillespie M, Haw R, Loney F, May B, Milacic M, Rothfels K, Sevilla C,
  Shamovsky V, Shorser S, Varusai T, Weiser J, Wu G, Stein L, Hermjakob H,
  D'Eustachio P (2020).
\newblock \enquote{The Reactome Pathway Knowledgebase.}
\newblock \emph{Nucleic Acids Research}, \textbf{48}(D1), D498--D503.
\newblock \doi{10.1093/nar/gkz1031}.

\bibitem[{J\"oreskog and S\"orbom(2018)}]{LISREL10}
J\"oreskog KG, S\"orbom D (2018).
\newblock \emph{\proglang{LISREL} 10 for Windows}.
\newblock Skokie, IL.
\newblock \urlprefix\url{https://www.ssicentral.com/index.php/products/lisrel}.

\bibitem[{Kanehisa and Goto(2000)}]{KEGG2000}
Kanehisa M, Goto S (2000).
\newblock \enquote{KEGG: Kyoto Encyclopedia of Genes and Genomes.}
\newblock \emph{Nucleic Acids Research}, \textbf{28}(1), 27--30.
\newblock \doi{10.1093/nar/28.1.27}.

\bibitem[{Kou \emph{et~al.}(1981)Kou, Markowsky, and Berman}]{kou1981}
Kou L, Markowsky G, Berman L (1981).
\newblock \enquote{A fast algorithm for Steiner trees.}
\newblock \emph{Acta Informatica}, \textbf{15}(2), 141--145.
\newblock \doi{10.1007/BF00288961}.

\bibitem[{Larson and Owen(2015)}]{Permutation2015}
Larson JL, Owen AB (2015).
\newblock \enquote{Moment based gene set tests.}
\newblock \emph{BMC Bioinformatics}, \textbf{16}, 132.
\newblock \doi{10.1186/s12859-015-0571-7}.

\bibitem[{Lefcheck(2016)}]{piecewiseSEM2016}
Lefcheck JS (2016).
\newblock \enquote{piecewiseSEM: Piecewise Structural Equation Modelling in R
  for Ecology, Evolution, and Systematics.}
\newblock \emph{International Journal of Epidemiology}, \textbf{7}(5),
  573--579.
\newblock \doi{10.1111/2041-210X.12512}.

\bibitem[{Li \emph{et~al.}(2015)Li, Chen, Sears, Gao, Klein, Karydas,
  Geschwind, Rosen, Boxer, Guo, Pellegrini, Horvath, Miller, Geschwind, and
  Coppola}]{FTD2015}
Li Y, Chen J, Sears R, Gao F, Klein E, Karydas A, Geschwind M, Rosen H, Boxer
  A, Guo W, Pellegrini M, Horvath S, Miller B, Geschwind D, Coppola G (2015).
\newblock \enquote{An epigenetic signature in peripheral blood associated with
  the haplotype on 17q21.31, a risk factor for neurodegenerative tauopathy.}
\newblock \emph{PLoS Genetics}, \textbf{10}(3), e1004211.
\newblock \doi{10.1523/JNEUROSCI.2939-14.2015}.

\bibitem[{Liu \emph{et~al.}(2020)Liu, Ma, Zhao, Nussinov, Zhang, Cheng, and
  Zhang}]{Network2020}
Liu C, Ma Y, Zhao J, Nussinov R, Zhang YC, Cheng F, Zhang ZK (2020).
\newblock \enquote{Computational Network Biology: Data, Models, and
  Applications.}
\newblock \emph{Physics Reports}, \textbf{846}, 1--66.
\newblock \doi{10.1016/j.physrep.2019.12.004}.

\bibitem[{Mahalanobis(1936)}]{mahalanobis1936}
Mahalanobis PC (1936).
\newblock \enquote{A General Approach to Confirmatory Factor Analysis.}
\newblock \emph{Proceedings of the National Institute of Sciences of India},
  \textbf{2}(1), 49--55.
\newblock
  \urlprefix\url{https://insa.nic.in/writereaddata/UpLoadedFiles/PINSA/Vol02_1936_1_Art05.pdf}.

\bibitem[{Marchetti \emph{et~al.}(2020)Marchetti, Drton, and Sadeghi}]{ggm2020}
Marchetti GM, Drton M, Sadeghi K (2020).
\newblock \emph{ggm: Graphical Markov Models with Mixed Graphs}.
\newblock R package version 2.5,
  \urlprefix\url{https://CRAN.R-project.org/package=ggm}.

\bibitem[{Massara \emph{et~al.}(2016)Massara, Matteo, and Aste}]{TMFG2016}
Massara GP, Matteo TD, Aste T (2016).
\newblock \enquote{Network Filtering for Big Data: Triangulated Maximally
  Filtered Graph.}
\newblock \emph{Journal of complex Networks}, \textbf{5}(2), 161--178.
\newblock \doi{10.1093/comnet/cnw015}.

\bibitem[{Muth\'en and Muth\'en(2017)}]{Mplus8}
Muth\'en LK, Muth\'en BO (2017).
\newblock \emph{Mplus User's Guide, Version 8}.
\newblock Los Angeles, CA.
\newblock \urlprefix\url{http://www.statmodel.com/}.

\bibitem[{Neumeyer \emph{et~al.}(2019)Neumeyer, Hemani, and Zeggini}]{eQTL2019}
Neumeyer S, Hemani G, Zeggini E (2019).
\newblock \enquote{Strengthening Causal Inference for Complex Disease Using
  Molecular Quantitative Trait Loci.}
\newblock \emph{Trends in Molecular Medicine}, \textbf{26}(2), 232--241.
\newblock \doi{10.1016/j.molmed.2019.10.004}.

\bibitem[{Newman and Girvan(2004)}]{ebc2004}
Newman MEJ, Girvan M (2004).
\newblock \enquote{Finding and evaluating community structure in networks.}
\newblock \emph{Physical Review E}, \textbf{69}(2), 026113.
\newblock \doi{10.1103/PhysRevE.69.026113}.

\bibitem[{Palluzzi \emph{et~al.}(2017)Palluzzi, Ferrari, Graziano, Novelli,
  Rossi, Galimberti, Rainero, Benussi, Nacmias, Bruni, Cusi, Salvi, Borroni,
  and Grassi}]{ItalianFTD2017}
Palluzzi F, Ferrari R, Graziano F, Novelli V, Rossi G, Galimberti D, Rainero I,
  Benussi L, Nacmias B, Bruni AC, Cusi D, Salvi E, Borroni B, Grassi M (2017).
\newblock \enquote{A novel network analysis approach reveals DNA damage,
  oxidative stress and calcium/cAMP homeostasis-associated biomarkers in
  frontotemporal dementia.}
\newblock \emph{PLoS One}, \textbf{12}(10), e0185797.
\newblock \doi{10.1371/journal.pone.0185797}.

\bibitem[{Pearl(1998)}]{SEM1998}
Pearl J (1998).
\newblock \enquote{Graphs, Causality, and Structural Equation Models.}
\newblock \emph{Sociological Methods \& Research}, \textbf{27}(2), 226--284.
\newblock \doi{10.1177/0049124198027002004}.

\bibitem[{Pearl(2009)}]{Causality2009}
Pearl J (2009).
\newblock \emph{Causality: Models, reasoning, and inference}.
\newblock 2nd edition. Cambridge University Press, New York, NY, USA.

\bibitem[{Pons and Latapy(2005)}]{wtc2005}
Pons P, Latapy M (2005).
\newblock \enquote{Computing communities in large networks using random walks
  (long version).}
\newblock \eprint{physics/0512106}.

\bibitem[{{R Core Team}(2020)}]{R2020}
{R Core Team} (2020).
\newblock \emph{R: A Language and Environment for Statistical Computing}.
\newblock R Foundation for Statistical Computing, Vienna, Austria.
\newblock \urlprefix\url{https://www.R-project.org/}.

\bibitem[{Reimand \emph{et~al.}(2019)Reimand, Isserlin, Voisin, Kucera,
  Tannus-Lopes, Rostamianfar, Wadi, Meyer, Wong, Xu, Merico, and
  Bader}]{enrichment2019}
Reimand J, Isserlin R, Voisin V, Kucera M, Tannus-Lopes C, Rostamianfar A, Wadi
  L, Meyer M, Wong J, Xu C, Merico D, Bader GD (2019).
\newblock \enquote{Pathway enrichment analysis and visualization of omics data
  using g:Profiler, GSEA, Cytoscape and EnrichmentMap.}
\newblock \emph{Nature Protocols}, \textbf{14}(1), 482--517.
\newblock \doi{10.1038/s41596-018-0103-9}.

\bibitem[{Ritchie \emph{et~al.}(2015)Ritchie, Holzinger, Li, Pendergrass, and
  Kim}]{Interactions2015}
Ritchie M, Holzinger E, Li R, Pendergrass S, Kim D (2015).
\newblock \enquote{Methods of Integrating Data to Uncover Genotype-Phenotype
  Interactions.}
\newblock \emph{Nature Review Genetics}, \textbf{16}(2), 85--97.
\newblock \doi{10.1038/nrg3868}.

\bibitem[{Rosseel(2012)}]{lavaan2012}
Rosseel Y (2012).
\newblock \enquote{lavaan: An R Package for Structural Equation Modeling.}
\newblock \emph{Journal of Statistical Software}, \textbf{48}(2), 1--36.
\newblock \doi{10.18637/jss.v048.i02}.

\bibitem[{Sch\"afer \emph{et~al.}(2017)Sch\"afer, Opgen-Rhein, Zuber,
  Ahdesmaki, Silva, and Strimmer}]{corpcor2017}
Sch\"afer J, Opgen-Rhein R, Zuber V, Ahdesmaki M, Silva APD, Strimmer K (2017).
\newblock \emph{corpcor: Efficient Estimation of Covariance and (Partial)
  Correlation}.
\newblock R package version 1.6.9,
  \urlprefix\url{https://CRAN.R-project.org/package=corpcor}.

\bibitem[{Sch\"afer and Strimmer(2005)}]{Shrinkage2005}
Sch\"afer J, Strimmer K (2005).
\newblock \enquote{A shrinkage approach to large-scale covariance matrix
  estimation and implications for functional genomics.}
\newblock \emph{Stat Appl Genet Mol Biol.}, \textbf{4}(1).

\bibitem[{Schermelleh-Engel and Moosbrugger(2003)}]{SEM2003}
Schermelleh-Engel K, Moosbrugger H (2003).
\newblock \enquote{Evaluating the fit of structural equation models: tests of
  significance and descriptive goodness-of-fit measures.}
\newblock \emph{Methods of Psychological Research Online}, \textbf{8}(2),
  23--74.
\newblock \urlprefix\url{http://www.mpr-online.de}.

\bibitem[{Shendure and Aiden(2012)}]{Sequencing2012}
Shendure J, Aiden EL (2012).
\newblock \enquote{The Expanding Scope of DNA Sequencing.}
\newblock \emph{Journal of Molecular Biology}, \textbf{30}(11), 1084--1094.
\newblock \doi{10.1038/nbt.2421}.

\bibitem[{Shipley(2000)}]{SEM2000}
Shipley B (2000).
\newblock \enquote{A new inferential test for path models based on DAGs.}
\newblock \emph{Structural Equation Modeling}, \textbf{7}(2), 206--218.
\newblock \doi{10.1207/S15328007SEM0702_4}.

\bibitem[{Shipley(2002)}]{BAP2002}
Shipley B (2002).
\newblock \enquote{Start and Stop Rules for Exploratory Path Analysis.}
\newblock \emph{Structural Equation Modeling A Multidisciplinary Journal},
  \textbf{9}(4), 554--561.
\newblock \doi{10.1207/S15328007SEM0904_5}.

\bibitem[{Shipley(2016)}]{Causality2016}
Shipley B (2016).
\newblock \emph{Cause and Correlation in Biology}.
\newblock 2st edition. Cambridge University Press, Cambridge, England, UK.
\newblock \doi{10.1017/CBO9781139979573}.

\bibitem[{Shojaie and Michailidis(2010)}]{DAG2010}
Shojaie A, Michailidis G (2010).
\newblock \enquote{Penalized Likelihood Methods for Estimation of Sparse
  High-Dimensional Directed Acyclic Graphs.}
\newblock \emph{Biometrika}, \textbf{97}(3), 519--538.
\newblock \doi{10.1093/biomet/asq038}.

\bibitem[{Spirtes \emph{et~al.}(2000)Spirtes, Glymour, and
  Scheines}]{Causality2000}
Spirtes P, Glymour CN, Scheines R (2000).
\newblock \emph{Causation, Prediction, and Search}.
\newblock 2nd edition. The MIT Press, Cambridge, MA, USA.

\bibitem[{Szklarczyk \emph{et~al.}(2019)Szklarczyk, Gable, Lyon, Junge, Wyder,
  Huerta-Cepas, Simonovic, Doncheva, Morris, Bork, Jensen, and von
  Mering}]{STRING2019}
Szklarczyk D, Gable AL, Lyon D, Junge A, Wyder S, Huerta-Cepas J, Simonovic M,
  Doncheva NT, Morris JH, Bork P, Jensen LJ, von Mering C (2019).
\newblock \enquote{STRING v11: Protein-Protein Association Networks With
  Increased Coverage, Supporting Functional Discovery in Genome-Wide
  Experimental Datasets.}
\newblock \emph{Nucleic Acids Research}, \textbf{47}(D1), D607--D613.
\newblock \doi{10.1093/nar/gky1131}.

\bibitem[{Textor \emph{et~al.}(2016)Textor, van~der Zander, Gilthorpe,
  Liskiewicz, and Ellison}]{dagitty2016}
Textor J, van~der Zander B, Gilthorpe M, Liskiewicz M, Ellison G (2016).
\newblock \enquote{Robust Causal Inference Using Directed Acyclic Graphs: the R
  Package `dagitty'.}
\newblock \emph{International Journal of Epidemiology}, \textbf{45}(6),
  1887--1894.
\newblock \doi{10.1093/ije/dyw341}.

\bibitem[{Tibshirani \emph{et~al.}(2012)Tibshirani, Bien, Friedman, Hastie,
  Simon, Taylor, and Tibshirani}]{glmnet2012}
Tibshirani R, Bien J, Friedman J, Hastie T, Simon N, Taylor J, Tibshirani RJ
  (2012).
\newblock \enquote{Strong rules for discarding predictors in lasso‐type
  problems.}
\newblock \emph{Royal Statistical Society: Series B (Statistical Methodology)},
  \textbf{74}(2), 245--266.
\newblock \doi{10.1111/j.1467-9868.2011.01004.x}.

\bibitem[{Verhulst \emph{et~al.}(2017)Verhulst, Maes, and Neale}]{GWSEM2017}
Verhulst B, Maes HH, Neale MC (2017).
\newblock \enquote{GW-SEM: A Statistical Package to Conduct Genome-Wide
  Structural Equation Modeling.}
\newblock \emph{Behavior Genetics}, \textbf{47}(3), 345--359.
\newblock \doi{10.1007/s10519-017-9842-6}.

\bibitem[{Verma and Pearl(1990)}]{Causality1990}
Verma T, Pearl J (1990).
\newblock \enquote{Causal Networks: Semantics and Expressiveness.}
\newblock \emph{Machine Intelligence and Pattern Recognition}, \textbf{9}(1),
  69--76.
\newblock \doi{10.1016/B978-0-444-88650-7.50011-1}.

\bibitem[{Wang and Zhao(2019)}]{cate2019}
Wang J, Zhao Q (2019).
\newblock \emph{cate: High Dimensional Factor Analysis and Confounder Adjusted
  Testing and Estimation}.
\newblock R package version 1.1,
  \urlprefix\url{https://CRAN.R-project.org/package=cate}.

\bibitem[{Williams(2020)}]{GGMncv2020}
Williams D (2020).
\newblock \emph{GGMncv: Gaussian Graphical Models with Non-Convex Penalties}.
\newblock R package version 1.1.0,
  \urlprefix\url{https://CRAN.R-project.org/package=GGMncv}.

\bibitem[{Witte \emph{et~al.}(2020)Witte, Henckel, Maathuis, and
  Didelez}]{oset2020}
Witte J, Henckel L, Maathuis MH, Didelez V (2020).
\newblock \enquote{On Efficient Adjustment in Causal Graphs.}
\newblock \emph{Journal of Machine Learning Research}, \textbf{21}(246), 1--45.
\newblock \urlprefix\url{http://jmlr.org/papers/v21/20-175.html}.

\bibitem[{Yu \emph{et~al.}(2015)Yu, Hillebrand, Tewarie, Meier, van Dijk,
  Mieghem, and Stam}]{MST2015}
Yu M, Hillebrand A, Tewarie P, Meier J, van Dijk B, Mieghem PV, Stam CJ (2015).
\newblock \enquote{Hierarchical clustering in minimum spanning trees.}
\newblock \emph{BMC Bioinformatics}, \textbf{25}, 023107.
\newblock \doi{10.1063/1.4908014}.

\bibitem[{Zhang \emph{et~al.}(2015)Zhang, Hamagami, Grimm, and
  McArdle}]{RAMpath2015}
Zhang Z, Hamagami F, Grimm KJ, McArdle JJ (2015).
\newblock \enquote{Using R Package RAMpath for Tracing SEM Path Diagrams and
  Conducting Complex Longitudinal Data Analysis.}
\newblock \emph{Structural Equation Modeling: A Multidisciplinary Journal},
  \textbf{22}(1), 132--147.
\newblock \doi{10.1080/10705511.2014.935257}.

\bibitem[{Zhao \emph{et~al.}(2012)Zhao, Liu, Roeder, Lafferty, and
  Wasserman}]{huge2012}
Zhao T, Liu H, Roeder K, Lafferty J, Wasserman L (2012).
\newblock \enquote{The huge Package for High-dimensional Undirected Graph
  Estimation in R.}
\newblock \emph{Journal of Machine Learning Research}, \textbf{13}(1),
  1059--1062.
\newblock
  \urlprefix\url{https://cran.r-project.org/web/packages/huge/vignettes/vignette.pdf}.

\bibitem[{Zhou(2009)}]{ThresholdLASSO2009}
Zhou S (2009).
\newblock \enquote{Thresholding Procedures for High Dimensional Variable
  Selection and Statistical Estimation.}
\newblock \emph{NIPS}, pp. 2304--2312.

\end{thebibliography}

\end{document}